\documentclass{aastex}
\usepackage[onecolumn]{emulateapj5}
\usepackage{graphicx}
\shorttitle{Two-Phase AGN Disk Wind Models}
\shortauthors{Everett}

\begin{document}

\title{Semianalytic Models of Two-Phase Disk Winds in Active Galactic
Nuclei with Combined Hydromagnetic and Radiative Driving}
\author{John E. Everett} 
\affil{Department of Astronomy and
Astrophysics, University of Chicago, 5640 S. Ellis Avenue, Chicago, IL
60637} 
\slugcomment{Submitted to ApJ}

\begin{abstract}
We present a semianalytic model of steady-state magnetically and
radiatively driven disk outflows in Active Galactic Nuclei (AGNs)
consisting of a continuous wind with embedded clouds.  The continuous
outflow is launched from the disk surface as a centrifugally driven
wind, whereas the clouds are uplifted from the disk by the ram
pressure of the continuous outflow.  The inner regions of the outflow
shield the outer portions from the strong ionizing central continuum,
enabling gas in the outer regions to be radiatively accelerated.  In
this paper, we describe the model in detail, outline the tests used to
verify its accuracy, and then compare it to other outflow models
already in use, showing that some previous AGN wind simulations do not
consider radiative transfer in the detail necessary for accurate
modeling. We also perform a comprehensive parameter study to explore
the dependence of the continuous wind properties on the relevant
physical parameters.  We find that the introduction of clouds has a
significant impact on the geometry, density, and terminal velocities
of the continuous wind: in particular, changes of up to 50\% in the
wind velocity can be directly attributed to the effect of clouds.
Within this survey, line driving is often the dominant acceleration
mechanism for clouds, owing to their small volume filling factor,
which yields an overall low cloud opacity.  Continuum driving can also
become important for dusty clouds.  For the continuous wind alone, we
find that radiative acceleration can have a significant impact on both
the terminal velocities and the kinematics near the surface of the
disk. However, unlike the clouds, the radiative acceleration of the
continuous wind comes largely from continuum opacities, not from line
driving as in some previous models.  Line driving produces comparable
acceleration only for parts of the wind far from the central source,
where the ionization state of the gas drops due to the lower incident
flux.  When the winds are dusty, their terminal velocities change by
approximately 20\% to 30\% (depending on the type of dust) over pure
centrifugally-driven outflows.
\end{abstract}

\section{Introduction}\label{Intro}

Active Galactic Nuclei (AGNs) often declare themselves boisterously to
observers: well-known for their incredibly bright nuclei, they can
also display emission line widths of up to several thousand ${\rm
km~s}^{-1}$, and absorption lines in the optical, UV, and X-ray, with
blueshifts up to 60,000 ${\rm km~s}^{-1}$.  They do not yet, however,
communicate as well with theorists: we lack a physical picture of the
geometry and kinematics of gas in the core of AGNs.  Researchers agree
that AGNs house supermassive black holes, but other components of the
theoretical picture spark heated debate.  The payoff for building a
successful model of the geometry and kinematics of gas in AGNs is very
high: not only will we gain understanding of how AGNs work, and
insight into the almost ubiquitous phenomenon of accretion and the
physics of black holes, but we may also learn about how galaxies
evolve \citep[e.g.,][]{Blandford2001}.

In the absence of any well agreed-upon theoretical model, researchers
have put together a phenomenological picture of the structure of AGN,
called the ``Unification Model''\citep{Ant93, UP95}.  As commonly
cited, this paradigm consists of a supermassive black hole with an
accretion disk, as well as a coplanar ``torus'' of gas and dust that
orbits near the outskirts of the accretion disk; the location,
composition, and dimensions of the torus remain unknown.  Gas,
commonly modeled as clouds and postulated to exist above the accretion
disk, produces both the broad emission lines (within the Broad
Emission Line Region, or BELR) and, further from the central source,
the narrow emission lines.  This model has proven itself useful
through its explanation of the difference between AGNs that contain
only narrow emission lines, and those that contain both broad and
narrow lines: if the AGN is oriented so that our line of sight
intersects the torus, it obscures the broad lines region, and we see
only narrow emission lines.  In recent years, models that further
elucidate the geometry and kinematics of gas in the unification model
have also been proposed \citep[e.g.,][]{Elvis00,Ganguly01}; although
these models cite the need for such mechanisms as radiative
acceleration, they remain chiefly empirical.

Observational hints are now appearing that the unified, ``inclination
angle'' paradigm for AGNs may not work in all objects.  In fact,
relatively few AGNs have been used as proof of this unification model
\citep{Axon01}.  In addition, inclination-angle independent measures
of various AGNs seem to show variations which are not in agreement
with the unified model \citep{Weymann02}.  Recent X-ray observations
of AGNs seem to show objects that have large X-ray absorbing columns,
and yet have broad optical emission lines, while other sources with
only narrow optical emission lines have no apparent X-ray absorption
\citep{Pappa01}.  Thus, the empirical models of gas distribution in
the central regions of AGNs might not be as widely applicable as
previously thought.  At the root of this difficulty lies the fact that
we are still quite ignorant of what forces dominate in different parts
of the nuclei of active galaxies, or what forces may dominate in
different AGNs.

To solve this problem, and develop consistent dynamical models of
AGNs, researchers have appealed to two major sources of acceleration:
magnetohydrodynamic driving of clouds \citep[e.g.,][]{Emmer92,Bot97},
and radiative acceleration of continuous winds
\citep[e.g.,][]{MCGV95,Pro00}.  One can also see in the above very
short summary that the field is somewhat divided on whether we are
watching continuous outflows or ``clumpy'' outflows [however, the
continuous wind simulations of \citet{Pro00} can produce condensations
that may be the source of a multiphase medium].  Both of these types
of models have had successes in modeling AGN observations; for
instance, both the magnetohydrodynamic model \citep{Bot97} and the
radiative acceleration model \citep{MC97} have reproduced the observed
single-peak broad emission lines observed in AGNs.  Since these models
have fit the basic observational traits of AGNs, there seems to be no
clear way to select which physics may be dominant in which AGNs, and
we are left with two very different outflow scenarios.  The challenge
remains, therefore, to develop tools that can distinguish between
these models and that will allow us to test for the presence of
individual forces or components.  A general model seems necessary: one
that self-consistently handles all of the elements that have been
previously proposed, so that we can systematically test one inclusive
model against the observations.  Such a model may help us gain insight
into how AGN winds work, how gas is distributed in the cores of AGNs,
and may aid in testing theories about accretion disks and AGN
evolution.

This paper presents such a tool, including all of the above mentioned
elements and their interactions (such as the interplay between
magnetic forces and radiative driving, and the drag forces between
the continuous wind and the clouds) to provide a physically
self-consistent ``platform'' from which to test all of the different
components, together.



This paper outlines our motivation for such a model, defines the
particular construction we have chosen, and introduces a parameter
survey to aid intuition of how the included processes interact.

The paper begins with an outline of past models in
\S\ref{modelOverview}, ending with an overview of our model.  In
\S\ref{detailedModel}, we describe the various components of our
model, and in \S\ref{Results} we present early results from our
calculations, showing the dependences of wind velocities and densities
on various input parameters.  We summarize our findings in
\S\ref{Conclusions}.

\section{AGN Outflow Models: The Current State of the Art}\label{modelOverview}

This section summarizes previously proposed models; this may be
useful, as our work builds on the results of past work, including
elements from many of the models mentioned.

\subsection{``Clumpy'' Winds}

Many of the models used today to reproduce AGN observations are
photoionization codes that invoke discrete clouds
\citep[e.g.,][]{Ham01, dK01}.  In many cases, no dynamical information
is input to the model; these winds are modeled as clumps of gas that
intercept the radiation from the central source \citep[see,
however,][]{Chelouche99}.  These cloud models have an extensive
history in fitting AGN winds \citep[see][]{Netzer90}: for instance,
they have been used to explain how both high and low ionization lines
occur together (using optically thick clouds) with similar line
profiles.  They are also attractive since such inhomogeneous winds
predict much smaller mass outflow rates compared to continuous winds.
In addition, continuous wind simulations seem unstable to producing
density inhomogeneities, making clouds a somewhat ubiquitous component
in AGN wind models.

However, there has been some doubt cast on this picture recently, as
high-resolution analysis of emission line profiles seems to suggest
many more clouds than specified by photoionization analysis
\citep{Arav97, Arav98, Dietrich99}.  As a result, some researchers
have turned to continuous wind models instead of discrete clouds.
Another problem with the cloud models is that these clouds must be
confined in some way in order to retain their distinct identities, and
finding a confining mechanism has remained a difficult task.  Some
researchers have called on magnetic confinement \citep{Rees87}, but
others claim that magnetic confinement has never been proven to work.
The response to this argument has typically been that since the solar
wind confines clouds within the magnetic structure of its outflow,
perhaps it is possible for an AGN outflow to do the same
\citep{Weymann02}.

\subsection{Radiative Acceleration of Continuous Winds}

As an alternative to clumpy winds, some researchers have developed
continuous wind models, usually powered by radiative acceleration.
Radiative acceleration is the force felt by atoms due to momentum
conservation: atoms de-excite isotropically after absorbing radially
(thus, anisotropically) streaming photons from the central source.
Since the late 1970's, radiative acceleration has been a popular
mechanism for explaining stellar winds
\citep[e.g.,][]{CAK75,Abbott78}, and has become an attractive
candidate for AGN winds \citep[e.g.,][]{Arav94, MCGV95, Pro00} because we know
that there is an intense radiation field in AGNs. Therefore, it seems
very likely that radiative acceleration should be important somewhere
within the outflow.  In addition, radiative acceleration calculations
have shown that the observed terminal outflow velocities can be
achieved \citep[hereafter, MCGV95]{MCGV95}.  Researchers have even
been able to reproduce the characteristic AGNs single-peaked broad
line profiles using radiatively accelerated, continuous winds
\citep{MC97}.  There have also been several observational hints that
radiative acceleration is an important force in AGN: for instance, the
dependence of absorber terminal velocity on AGN luminosity
\citep{LaorBrandt02} and the phenomenon of spectral absorption ``line
locking'' \citep{Arav96}.

However, radiative acceleration becomes important only with the right
balance of UV and X-ray radiation: too much flux in the X-ray band can
strip atoms of their electrons, leaving many fewer line transitions
for atoms to absorb and accelerate with \citep{Pro02a}.  But if one
decreases the continuum enough to leave the atoms intact, the UV
radiation absorbed by the atomic transitions also decreases.  In order
to deliver the right amount of radiation to their winds, MCGV95 call
upon a modified MF87 spectrum in the X-ray band to mirror radio quiet
quasars: whereas \citet[hereafter MF87]{MF87} has an $\alpha_{\rm OX}$
of 1.4 and a power-law slope of -0.7 above 0.4 keV, MCGV95 call upon
an $\alpha_{\rm OX}$ of 1.5 and a power-law slope of -1 throughout the
X-ray band.  ($\alpha_{\rm OX}$ is the spectral index that describes
the power law slope connecting the continuum at 2500 \AA and 2 keV.)
MCGV95 also utilize a gas shield to attenuate that continuum.  As this
shield is crucial to their model, a self-consistent explanation for
the presence of that shield would be a very important element in any
AGN model.


\subsection{Magnetic Acceleration of Continuous Winds or Clouds}

The use of magnetic stresses as a possible source of acceleration and
collimation of AGN winds stretches back to \citet[hereafter
BP82]{BP82}, and since then has been developed by other researchers
\citep[e.g.,][]{Emmer92, KK94, Bot97, KKE99}.  The primary reason that
such magnetic field models are attractive is that not only could they
help form the collimated jets that we observe at large scales, but
more importantly, they transport angular momentum from the accretion
disk, allowing matter to fall towards the central source.  The classic
description of this model involves visualizing the magnetic field
lines as wires, and gas elements as beads on that wire (assuming that
the gas and the field are well-coupled, which is a good approximation
if the gas is well ionized).  If the magnetic field meets the disk at
an angle $\leq 30^{\circ}$ to the vertical, then due to centrifugal
forces, the gas ``sees'' a potential drop along the magnetic field
line.  Because of this potential drop, the gas launches from the
accretion disk, moving along the field lines.  Thus a wind moves
upward from the disk, and will later be collimated by the same
magnetic field lines.  This collimation occurs due to the increase in
importance of the matter's inertia relative to the magnetic field
strength: as the wind rises, the magnetic field amplitude drops, and
the rotational motion of the gas (retained from the disk) starts to
dominate the gas streamlines, twisting the magnetic field into a
coil-like shape that confines the outflow.  The self-contained
launching and collimation mechanism in this theory along with its
possible role in helping matter accrete make this one of the most
appealing, generally-applicable models to a wide variety of disk/wind
systems \citep{Livio99}.

	In addition to explaining collimated jets and helping matter
accrete in the disk, magnetic winds may be valuable for AGN models in
that they can also supply magnetic pressure and confine clouds
embedded in the wind.  Also, radiatively accelerated wind models will
have difficulty launching a wind in heavily-shielded (or otherwise
very low luminosity) regions of the outflow, which could be the case
near the surface of the accretion disk.  Magnetic stresses could lift
the wind up to a starting point where the radiative acceleration would
become effective, regardless of the shielding or the luminosity of the
AGN (a very luminous accretion disk could also play an important role
in wind launching, however).  Finally, magnetic winds most easily
accelerate gas that is highly ionized and could thus launch an
unshielded wind; in comparison, radiative acceleration drops when the
gas is highly ionized since such gas would not have a large array of
lines to absorb the photons necessary to accelerate the gas.  The
magnetically launched wind could then provide a gas shield for winds
farther removed from the central source.

Magnetic wind models are also not without detractors: some researchers
argue that the required, relatively ordered magnetic field structure
does not exist, and that magnetic pressure is not adequate to confine
gas clouds (e.g., MCGV95).  Others argue that magnetic fields
represent ``extra forces'' that are not yet required in the face of
radiative acceleration models \citep{Weymann02}.

In addition, past models of magnetically accelerated winds have not
fully addressed the difficulties of mixing a magnetocentrifugal wind
with radiative acceleration, or with combining a continuous wind with
embedded clouds.  \citet[hereafter KK94]{KK94} contained an
approximate, schematic formula for radiative acceleration due to dust;
they did not attempt to include a detailed photoionization treatment,
and did not self-consistently include the radiative acceleration
within the framework of the self-similar MHD wind model.  Also,
\citet{Emmer92} analyzed AGN cloud dynamics by using a continuous wind
model in which the magnetic field structure was fixed.  Very little
previous work has been done on a model that includes both a continuous
wind and explicit clouds, the interaction between them (pressure
equilibrium and ram pressure), or the more large-scale problem of the
self-consistent solution of the magnetic and radiative acceleration
forces \citep[see, however, ][]{Pro02b}.

Each of the above models, with their respective geometries and forces,
have duplicated some aspect of AGN observations: some have matched
observed column densities, and in the case of dynamical models, have
even reproduced AGN emission line profiles \citep{Bot97,MC97}.  So
far, the different dynamical models seem to give very similar
observational predictions since they both result in somewhat similar
kinematics.  This has led to a serious difficulty: there is no easy
way to discriminate between these possibilities and learn what is
really powering AGN winds.  It may be important, therefore, to develop
models that can take these various forces into account and determine
what differences we can observe to discriminate among them, or to
determine the interplay of those forces within any given AGN.

\subsection{Self-Consistent Integrative Models}\label{ourModel}

We have developed a self-consistent model of Active Galactic
Nuclei winds, integrating all of the above forces within the same
program, and also taking account of their interactions.  This
semianalytic model includes magnetic acceleration and radiative
acceleration of both a continuous wind and clouds, as well as the drag
forces between the clouds and the wind.  In addition, through the use
of the photoionization simulation program Cloudy \citep[version 96
Beta 4;][]{F01}, we have included a detailed treatment of radiative
transfer.
        
\begin{figure}[ht]
\begin{center}
\includegraphics[0.75in,3.5in][8.5in,7.5in]{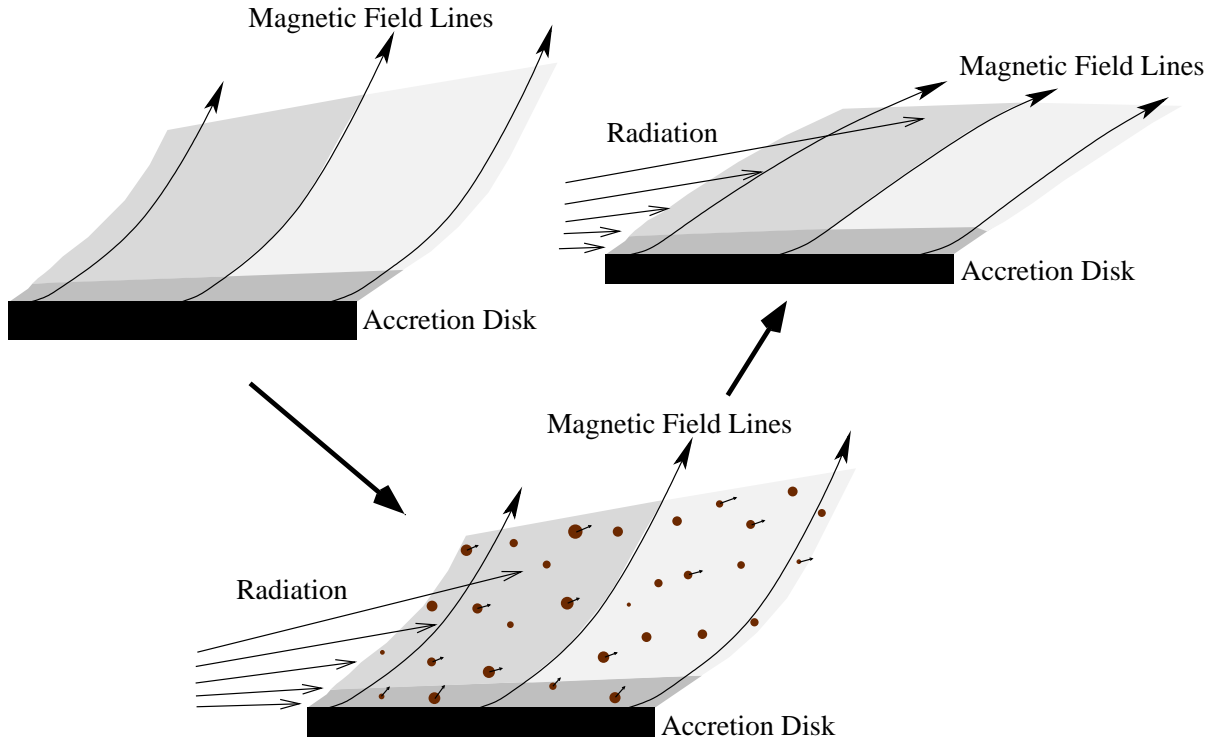} 
\end{center}
\caption{A schematic of the iterative scheme of our AGN wind model.
We represent a single iteration of the model, moving counter-clockwise
around the above diagram.  The model starts by solving the
self-similar MHD equations for the structure of a centrifugally driven
wind ({\it upper left}).  Next, we use photoionization simulations to
determine the ionization state of the gas, and then we use those
results to calculate the radiative acceleration of the wind.  We also
add clouds to the wind: clouds which are magnetically confined and
radiatively accelerated, and interact with the wind through drag
forces only ({\it lower center}).  We then apply the output from the
radiative acceleration calculation to the self-similar MHD wind, using
radiative acceleration to decrease gravity in a height-dependent way.
This modifies the wind's structure, taking into account the radiative
forces ({\it upper right}).  With the arrows, we denote the velocity
vectors of individual clouds; near the disk, where the shielding
column to the central source is high, the clouds feel predominantly
the drag of the wind, and are launched from the disk.  Further up in
the wind, the column of gas in the wind decreases as the wind
accelerates, the clouds intercept more radiation, and radiative
acceleration becomes the dominant force.\label{flowchart}}
\end{figure}

A schematic flow chart (see Fig.~\ref{flowchart}) gives a quick
overview of the model procedure: there are seven different programs at
work within that structure.  We start with a self-similar
magnetohydrodynamic model that gives the pure MHD wind solution, then
simulate the photoionization balance of that wind, and use the
resultant ionization balance to derive the radiative acceleration of
the wind and the embedded clouds.  Next, we input the radiative
acceleration of the wind back into the self-similar
magnetohydrodynamic model, modifying the structure of the wind.  We
also input into the self-similar model the drag forces felt by the
continuous wind due to the motion of the embedded clouds.  We then
repeat the process again, simulating the photoionization of that
modified wind and recalculating the radiative acceleration terms.  We
typically iterate three to five times to converge to a final
equilibrium solution.


Since we ``initialize'' the wind with a self-similar
magnetohydrodynamic flow, we are building in assumptions similar to
that of previous MHD models \citep[e.g.,][]{Emmer92}.  The assumption
of self-similarity is critical in enabling us to model the
magnetocentrifugal wind.  However, at the core of the self-similar
model is the assumption that variables scale via proscribed
power-laws; for instance, the spherical radial velocities all must
scale as $R^{-1/2}$, where $R$ is the spherical radial coordinate.  We
recognize this assumption by limiting our model to be strictly local,
i.e., the calculation is carried out and applicable only for a limited
range of launching radii $\Delta r \ll r_{\rm wind}$.  This limit must
also be imposed because of the extreme sensitivity of radiative
acceleration to ionization conditions, which means that the
calculations must by nature be local to where the photoionization
results are obtained.  Such local wind models seem, however, very
apropos to current AGN theories: researchers have postulated that the
wind is indeed outflowing in small sectors
\citep[e.g.,][]{Emmer92,dKB95,Arav98}, and such outflows may have even
been observed \citep{Arav96}.  This is interesting as such an outflow
does not require the large scale, ordered magnetic fields that may be
difficult for an AGN accretion disk to produce.  It is important,
however, to keep this limitation in mind: this model and its results
apply to local sectors, or flux tubes, of the outflowing wind.

Now that we have described the basics of our model, it may be
instructive to compare and contrast it to the recent wind model
examined in \citet{Pro02b}, where the combination of magnetic and
radiative forces in disk winds is also investigated.  \citet{Pro02b}
concentrates on numerical simulations of time-dependent winds with
line driving and magnetic forces.  These numerical simulations allow
large-scale models of outflows that are very valuable in understanding
global wind structures in many different astrophysical contexts.  In
contrast, the wind solutions presented here are much more localized,
as mentioned above.  However, as we will find in \S\ref{radTests},
this limitation is very important to accurately calculating the
radiation force on the wind, since we employ detailed photoionization
simulations.  We have chosen our model setup for its accuracy and
flexibility in radiative acceleration modeling; using Cloudy enables
computations not only of line driving but continuum driving, and
allows us the freedom to include dust and easily vary the incident
spectrum.  In addition, it is important to realize that the magnetic
winds produced in \citet{Pro02b} are not magnetocentrifugal (BP82)
outflows, as are the semianalytic winds that we present.  Further, our
models are steady-state, and not time-dependent.  We also explicitly
include confined clouds, while \citet{Pro02b} only simulates
continuous winds.  Finally, semi-analytic, steady-state models allow
an exploration of general behaviors through many parameter variations;
large-scale numerical simulations can usually vary only a few
parameters.  In summary, these models cover different facets of the
disk-wind problem, yielding valuable, different perspectives on a
complicated system.

\section{The Two-Phase Hydromagnetic and Radiative Wind Model}\label{detailedModel}

In this section, we describe all of the components in our model, and
outline the derivation of key equations.

\subsection{Magneto-Centrifugal Self-Similar Wind Solution}

The first part of our model solves the equations of motion for gas in
a magnetohydrodynamic (MHD), centrifugally-driven wind that is
uplifted, guided, and collimated by a magnetic field.

\subsubsection{Continuous Wind Equations}

To derive the equations governing the continuous wind, we start with
the equations of a stationary, axisymmetric magnetohydrodynamic flow
in cylindrical coordinates, and utilize the continuity equation,
conservation of angular momentum along the flow, and both the radial
and vertical momentum equations (much as in BP82 and KK94).  We
neglect thermal effects in the wind, therefore effectively assuming
that the wind starts out supersonic.  In deriving the equations of
motion, we use the same simplifications as BP82, except for the added
complication that energy is not conserved in our system, due to the
constant input of radiative energy into the system.  In the original
formation of BP82, conservation of energy supplied an additional
equation which allowed a simplification of their equations of motion
to two first order differential equations.  Because we lack an energy
constraint, we must integrate the equivalent of three first-order
differential equations, solving for three parameters simultaneously
instead of the two that are solved for in the case of BP82.  The
detailed setup and derivation of our equations of motion is are given
in Appendix~\ref{selfSimWindEqns}.

We start the integration of the momentum equations by specifying the
following initial parameters: the mass loading of the wind (the ratio
of mass flux to magnetic flux in the MHD wind, $\kappa \propto \frac{4
\pi \rho v_p}{B_p}$, where $\rho$ is the mass density of the wind,
$v_p$ is the poloidal velocity of the wind, and $B_p$ is the poloidal
magnetic field strength), the specific angular momentum of gas in the
wind, and the power-law exponent that describes the change in density
with spherical radius, $b$: $\rho \propto R^{-b}$.  We also input, as
parameters, the mass of the central black hole, $M_{\bullet}$, the
wind's launch radius on the disk, $r_0$, and the density at the base
of the wind (at $r_0$), $n_0$.  The program employs a ``shooting''
algorithm \citep[using the SLATEC routine DNSQ from][]{Powell70} to
integrate from the critical point (the Alfv\'en point) to the disk,
solving for the height of the critical point above the disk ($\chi_A$)
and the slope of the streamline at both the disk and the critical
point ($\xi'_0$ and $\xi'_A$) by matching the integration results to
boundary conditions on the disk.  After solving for the position of
the Alfv\'en point, we integrate the equations of motion from the disk
to a user-specified height beyond the Alfv\'en point, calculating the
run of velocity, density, and magnetic field along the gas flowlines.


\subsubsection{Testing}

        We have tested this code (without radiative acceleration)
against the results given in BP82 and have duplicated their results to
within 8\%.  This is fairly close to the previously reported 4\%
variance in recalculating their models \citep{S93}.

\subsection{Photoionization Simulations of the Wind}

We next use the photoionization code Cloudy \citep{F01} to simulate the
ionization state of the above MHD wind due to radiation from the
central AGN source.  We use photoionization simulations both of the
continuous wind and the clouds in our model to accurately calculate
the radiative acceleration of both components.  We can use any central
spectrum we wish, although we currently call on the MF87 and
MCGV95 spectra for the tests in this paper.

\subsubsection{Photoionization of the Continuous Wind}

Our goal in using Cloudy is to simulate the ionization state of the
gas and the radiative acceleration along a streamline within the wind.
We specify which streamline by giving Cloudy the distance of the base
of that streamline from the central continuum source, as well as the
wind's column density between the launch point and the innermost
region of the wind.  That part of the wind (between the central source
and the streamline of interest) acts as a shield for the streamline
under consideration.


However, there is a complication to running photoionization
simulations for these wind models: in many cases, near the base of the
wind, the gas could be optically thick.  Programs like Cloudy have an
understandably difficult time working with optically thick models, so
when we are in a region of the wind where the shielding column is
large enough to yield $\tau_{es} > 1$, we divide (computationally) the
wind into a $\tau_{es} \sim 1$ section and a remaining section that we
simulate as a pure electron scattering zone.  To do this, we compute
$\tau_{es}$ as follows:
\begin{eqnarray}
\tau_{es} & = & \sigma_T N_H = \sigma_T \int_{R_{in}}^{R_{out}} n(R) dR,
\end{eqnarray}
where $\sigma_T$ is the Thomson cross section, and $R$ is the
spherical, radial distance.  For the time being, we assume the same
scalings as BP82, and set $b = \frac{3}{2}$; other scalings are
possible \citep{CL94}.  We then write
\begin{eqnarray}
n(R) & = & n_0 \left( \frac{R_0}{R}  \right)^\frac{3}{2}. 
\end{eqnarray}
Substituting that into our expression for $\tau_{es}$, we find
\begin{eqnarray}
\tau_{es} & = & 2 \sigma_T n_0 R_0 \left( 1 - \frac{R_{out}}{R_0}
\right)^{\frac{1}{2}}.  
\end{eqnarray}

Again, if this $\tau_{es} > 1$, we can only simulate a $\tau_{es} \sim
1$ region in Cloudy, so we solve for the size of the column of gas in
front of our streamline that supplies $\tau_{es} = 1$, and pass that
part of the wind to Cloudy to simulate.  To include the effect of the
wind that we cannot have Cloudy simulate (that part of it closer to
the central source), we simply attenuate the AGN spectrum by
$e^{\tau-1}$ and pass that spectrum to Cloudy for the start of its
calculation.  When this program has finished, we record the ionization
state of the gas as well as the radiation field at that point in the
wind.

\subsubsection{Photoionization of the Clouds}

After simulating the radiative transfer through the wind, to our
streamline of interest, we run another set of photoionization
simulations to determine the photoionization state of clouds locally
in that wind.  But there is a complication here: the level of wind
shielding in some of our calculations is so high very near the disk
that Cloudy cannot simulate the ionization state there.  This becomes
a complication in the cloud models where we need to have some
estimates of the clouds' properties in order to understand their
launching. The details of how we define our initial cloud parameters
are given in Appendix~\ref{cloudInit}.

To keep the cloud parameters consistent with magnetic confinement, the
program iterates the Cloudy simulations of the clouds, adjusting the
density of the clouds until pressure balance between the clouds and
the magnetic pressure of the outside continuous wind is reached,
satisfying: $\frac{B_{wind}^2}{8 \pi} = n_{cloud} k T_{cloud}$ (the
code stops when they are equal to within 5\%).  When these Cloudy runs
are finished, the code outputs the density, ionization structure, and
radiation field in the clouds at each point in the wind.  With these
physical parameters set, we can calculate the radiative acceleration
of the clouds and the wind.

\subsection{Radiative Acceleration Calculations} 

At this point, we engage a FORTRAN program that uses Cloudy's results
for the ionization structure and radiation field to calculate the
radiative acceleration felt by the wind and clouds.  There are two
different kinds of radiative acceleration to consider: continuum
acceleration (which includes radiative acceleration on dust) and line
acceleration.  When we specify the radiative acceleration, it is
convenient to refer to $\Gamma(\theta)$
\begin{eqnarray}
\Gamma(\theta) & \equiv & \frac{a_{radiative}(\theta)}{g} \label{gammaDef}, 
\end{eqnarray}
where $a_{radiative}$ is the acceleration due to radiative forces, and
$g$ is the local gravitational acceleration.

\subsubsection{Line and Continuum Acceleration}

In general, for continuum and line acceleration, the radiative
acceleration is given by
\begin{eqnarray}
\Gamma & = &  \frac{\frac{n_e \sigma_T F}{\rho c} ( M_{cont} +
M_{lines})}{\frac{G M_{\bullet}}{r^2 + z^2}},
\end{eqnarray}
where $F$ is the incident flux, $n_e$ is the electron density, $\rho$
is the gas density, $c$ is the speed of light, $G$ is the
gravitational constant, $M_{\bullet}$ is the mass of the central black
hole, $r$ and $z$ are the radial and vertical distance from the black
hole, and $M_{lines}$ \& $M_{cont}$ are the ``force multipliers'' that
relate how much the radiative forces on the gas exceed the radiative
forces on electrons alone.  They are given below in terms of the
continuum opacity, $\chi_{\nu}$, and the line opacity, $\chi_l$
\begin{eqnarray}
M_{cont} & = & \frac{1}{n_e \sigma_T F} \int \chi_{\nu} F_{\nu} d\nu,
\\
M_{lines} & = & \frac{1}{F} \sum_l F_l \Delta\nu_l \frac{1 -
e^{-\eta_l t}}{t} \label{mLinesDef}, 
\end{eqnarray}
with
\begin{equation}
\eta_l \equiv  \frac{\chi_l}{\sigma_T n_e} \hspace{1in}
t \equiv \frac{\sigma_T n_e v_{th}}{\frac{dv_R}{dR}},
\end{equation}
where $\nu$ is the frequency, $F_l$ is the flux in the line at the
frequency of line $l$, $v_{th}$ is the thermal velocity in the
acceleration gas, $\Delta \nu_l = \nu v_{th}/c$ is the thermal line
width, and $\eta_l$ compares the opacity of the line (given the
ionization state of the gas) to the electron opacity, including all of
the atomic physics in the radiative acceleration calculation.  The
variable $t$ is often called the ``effective electron optical depth.''
This parameter sets line driving apart from all other radiative
transfer problems: $t$ encodes the dynamical information of the wind
in the radiative acceleration calculation.  In standard, stationary
radiative transfer problems, one calculates the optical depth in
atomic lines based purely on the linear depth and density of the
matter.  In an accelerating medium, one must also account for the
Doppler effect: a certain distance from the position of a line
photon's creation, the gas has accelerated to such an extent that, in
the rest frame of that downstream gas, the photon's energy has shifted
beyond the energy range that can be absorbed by that atomic line.
This length is termed the ``Sobolev length,'' and is a key parameter
in understanding the dynamics of radiatively driven flows
\citep{Sobolev58, Sobolev60, Castor70, CAK75, MM99}.  The parameter
$t$ includes the effect of the limiting Sobolev length by multiplying
the standard electron optical depth by $v_{th}/\frac{dv_R}{dR}$, which
is the length at which a photon has redshifted out of the thermal
width of a given line.

We calculate the force multipliers for every point in the wind where
we have photoionization simulation results, using the resonance lines
given in \citet{Verner96}.  Since $M_{cont}$ depends only on the
ionization state, as soon as we have photoionization data, we know the
continuum force multiplier.  We compute $M_{line}$ for a range of
values of the parameter $t$.  In the subsequent radiative acceleration
calculation, we linearly interpolate both of these tables to evaluate
the radiative acceleration, given the local ionization state and the
local velocity gradient.

\subsubsection{Dust Acceleration}\label{dustExpl}

To include radiative acceleration due to dust, we simply specify to
Cloudy that the gas includes dust (specifying to Cloudy both the type
of dust and the gas abundances corresponding to that dust).  In our
models so far, we have primarily used ISM gas abundances and dust
\citep{MRN77,DL84,MR91}.  We have also found wind solutions for clouds
with Orion abundances and dust \citep[the Orion dust differs in the
size cutoffs in the grain distribution: whereas the ISM dust has
grains of minimum size $a_{\rm min} = 0.0025~\mu{\rm m}$, the Orion
dust distribution has $a_{\rm min} = 0.03~\mu{\rm m}$, believed
appropriate to a UV-irradiated medium;][]{Baldwin91}.  For the grain
type specified, Cloudy then automatically includes the relevant dust
opacity in the continuum opacity that we use to calculate the
continuum force multiplier.

We include dust in the clouds by default to not only be consistent
with previous findings \citep{EKA02}, but also because the surface of
the accretion disk is often envisioned as having a high-temperature
corona: in such a corona, dust would be much more likely to survive in
the lower-temperature environment of the denser clouds.  However, for
full generality, the program also allows the user to add dust to the
continuous wind to observe the effects of that dust on the outflow.
Our parameter survey (see \S\ref{paramSurvey}) includes such models.

Within our model, we are careful to include dust only where the
photoionization models allow it to survive.  Primarily, this means
considering sublimation: for the ISM and Orion dust typically used in
these simulations, graphite and silicate dust sublimate at $\sim$1750
K and $\sim$1400 K, respectively.  We handle the process of dust
sublimation by including dust, automatically, where-ever the user
requests it, in our first photoionization simulations; if the dust
temperatures rise above the given sublimation temperatures, the
program removes dust from the calculation, reruns Cloudy, and does not
include dust in any subsequent photoionization simulations downstream
in the wind.

\subsubsection{Testing}

\begin{figure}[ht]
\begin{center}
\includegraphics[width=5in,angle=-90]{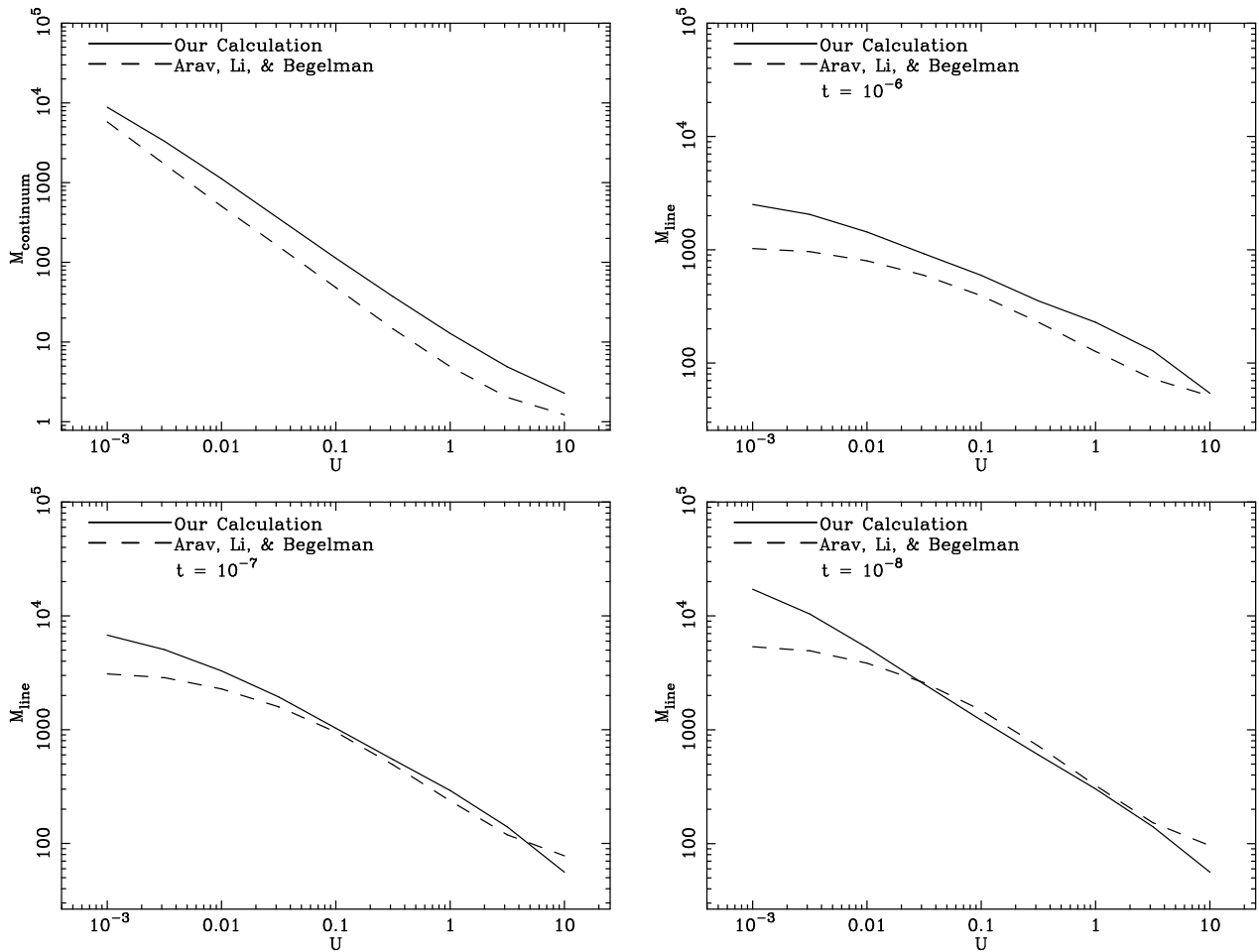} 
\end{center}
\caption{A comparison between the force multiplier calculation in our
code (solid lines) and the fits of \citet{Arav94} (dashed lines).  The
plot in the upper left is the comparison between our continuum force
multiplier, as a function of the ionization parameter, $U$, and the
earlier published fit.  The next three panels show a comparison of the
line force multiplier (which is also a function of $t$, besides $U$)
as functions of $U$ for three different values of $t$, given at the
top of each panel.  Overall, the agreement is fairly good, especially
since the fits were known to deviate at low ionization parameter, $U$.
The offset in the continuum multipliers is due to the different list
of continuum opacities used, but the similar shape is encouraging.
\label{forceMultCompare}}
\end{figure}

We have tested the computation of the force multipliers against
\citet{Arav94}, who also calculated the radiative acceleration from
photoionization simulations.  Figure~\ref{forceMultCompare} compares
our results against their fits (noting that there is a typo in their
eq. 2.9; Z.-Y. Li, personal communication), where we present the
radiation force multipliers as a function of the ionization parameter
U ($U$ is the ratio of hydrogen-ionizing photon density to hydrogen
number density $n$, given by $U \equiv Q/4{\pi}nR^2c$, where $Q$ is
the number of incident hydrogen-ionizing photons per second, and $R$
is the distance from the continuum source).  Overall, we find good
agreement, especially considering that \citet{Arav94} point out that
their fit deviates from rigorous calculations at low values of $U$ .
The increase of our continuum force multiplier over theirs is most
likely due to the different continuum opacity database included in
Cloudy 96 Beta 4 compared to the code that was used in \citet{Arav94}.
The multiplier values and trends with ionization parameter are still
clearly very similar, however, which is encouraging.

We have also tested that the sputtering timescale of dust within the
clouds and wind is less than the transit time of the gas through
regions of high sputtering rate.  To check this, we use the results of
\citet{Tielens94} to calculate the sputtering times (the times
required to completely sputter away grains) throughout the wind and
clouds.  The transit timescale exceeds the sputtering timescale for
the smallest ($a_{\rm min} = 50~$\AA) grains for all of the models
which include dust in the continuous wind; hence, sputtering is
important for these grains.  The clouds, on the other hand, have dust
with sputtering timescales $\geq 10^7$~years.  However, for the models
with very low-density continuous wind ($n_0 = 10^6~{\rm cm}^{-3}$),
the sputtering timescale in the clouds can be as short as $\sim
3000$~years, which is very close to the transit timescale of $\sim
1000$~years.  For the time being, we do not change the dust
distribution to take account of this sputtering; we simply note that a
significant number of small grains may be sputtered in the wind (and
perhaps some in the clouds).  These calculations have, however,
prompted us to examine solutions with an Orion dust distribution in
the continuous wind: Orion dust does not have the smallest sized
grains that are present in the ISM dust distribution.  The comparison
of Orion and ISM dust wind models will presented in
Fig.~\ref{WindVaryDust}.

\subsection{Integrating the Equation of Motion for the Wind}

Given the ionization and radiative acceleration information, our next
step is to find the radiative acceleration of the continuous wind.  We
take the MHD wind model already computed and calculate the additional
acceleration due to radiative forces.  To do this, we derive the
equation of motion for the wind, given by the Euler equation.  We can
then integrate that equation along the streamline of the MHD wind and
record $\Gamma(\theta)$, the acceleration due to radiative forces.

In its simplest form, Euler's Equation is given as
\begin{eqnarray}
\rho \left( \frac{\partial \mathbf{v}}{\partial t} + (\mathbf{v} \cdot
\nabla)\mathbf{v} \right) = \sum \mathbf{F}_i,
\end{eqnarray}
where $\mathbf{F}_i$ represent all of the different forces in our
system.

To specialize this equation for our wind, we first note that we are
examining steady-state systems, so we neglect $\frac{\partial
\mathbf{v}}{\partial t}$.  As already mentioned, we neglect the
thermal pressure but include gravitational, radiation, and Lorentz
forces.  In the expression below, we split the magnetic force term
into pressure and tension components.
\begin{eqnarray}
\rho (\mathbf{v} \cdot \nabla)\mathbf{v} & = & - [1 - \Gamma(\theta)]
\frac{G M \rho}{R^2}\hat{R} - \frac{1}{8 \pi} \nabla B^2 + \frac{1}{4
\pi} (\mathbf{B} \cdot \nabla)\mathbf{B}
\end{eqnarray}

For our calculations, it will be more intuitive to integrate the
equation of motion along the flow already given by the
magnetocentrifugal wind solution.  Thus, we take the dot product of
Euler's equation with $\hat{s}$, which we define as the direction
along the flow, expand, and simplify the left-hand side of the
equation to read
\begin{eqnarray}
((\mathbf{v} \cdot \nabla)\mathbf{v}) \cdot \hat{s} & = & v_p 
\frac{\partial v_p }{\partial s} -
\frac{v_{\phi}^2}{r} \sin \theta_F,
\end{eqnarray}
where we define 
\begin{eqnarray}
\theta_F & \equiv & \tan^{-1} \left( \frac{dr}{dz} \right).
\end{eqnarray}
In the same way, if we define 
\begin{eqnarray}
\theta \equiv \tan^{-1} \left( \frac{r}{z} \right),
\end{eqnarray}
we can write out the gravitational term as
\begin{eqnarray}
- [1 - \Gamma(\theta)] \frac{G M}{R^2}\hat{R} \cdot \hat{s} & = - [1 - \Gamma(\theta)] \frac{G M}{(r^2 + z^2)} \cos (\theta - \theta_F).
\end{eqnarray}
Next, we take the dot product of $\hat{s}$ with the magnetic terms to
find
\begin{eqnarray}
\left[ -\frac{1}{8 \pi \rho} \nabla B^2 + \frac{1}{4 \pi \rho}
(\mathbf{B} \cdot \nabla)\mathbf{B} \right] \cdot \hat{s} & = &
-\frac{B_{\phi}}{4 \pi \rho r} \frac{\partial (r B_{\phi})}{\partial s}.
\end{eqnarray}
Bringing all of those terms back together into the Euler Equation, we
can write
\begin{eqnarray}
v_p \frac{\partial v_p }{\partial s} - \frac{v_{\phi}^2}{r} \sin
\theta_F  & = & - [1 - \Gamma(\theta)] \frac{G M}{(r^2 + z^2)} \cos
(\theta - \theta_F) -\frac{B_{\phi}}{4 \pi \rho r} \frac{\partial (r
B_{\phi})}{\partial s}.
\end{eqnarray}

We still have a $v_{\phi}$ dependence, however, that we can eliminate
by appealing to the connection between $v_p$ and $v_{\phi}$ in the
Blandford \& Payne model, which gives
\begin{eqnarray}
v_{\phi} & = & \frac{v_p B_{\phi}}{B_p} + \Omega r.
\end{eqnarray}

We substitute this expression into our Euler Equation, which yields
\begin{eqnarray}
v_p \frac{\partial v_p }{\partial s} - \left(\frac{v_p B_{\phi}}{B_p}
+ \Omega r \right)^2 \frac{\sin \theta_F}{r} & = & - [1 -
\Gamma(\theta)] \frac{G M}{(r^2 + z^2)} \cos (\theta - \theta_F)
-\frac{B_{\phi}}{4 \pi \rho r} \frac{\partial (r B_{\phi})}{\partial s}.
\end{eqnarray}

	The only difficulty left is that, when we evaluate the
radiative acceleration parameter $t$, we need to know
$\frac{dv_R}{dR}$, the spherical, radial gradient of the spherical,
radial velocity.  Of course, we are integrating along the flow, and
can only approximate the velocity gradients perpendicular to the flow.
We approximate this gradient by calculating
\begin{eqnarray}
\frac{dv_R}{dR} \approx \frac{dv_R}{ds} \frac{ds}{dR}, \\
\frac{dv_R}{dR} \approx \cos^{2}(\theta - \theta_F) \frac{dv_p}{ds},
\end{eqnarray}
where the $\cos^{2}(\theta - \theta_F)$ term comes from approximating
the geometry of our outflow.

\subsubsection{Tests}

        We have tested the wind integration with radiative
acceleration turned off, where it reproduces the original self-similar
velocity profile to within one part in $10^{5}$.  With the radiative
acceleration turned on, and without clouds present, the entire code
has repeatedly converged within a few iterations to a ``puffed out''
magnetic wind structure, showing that radiation pressure does effect
the wind.  We have also tested this part of the code (again, without
clouds) against \citet{Pro00} and reproduced a similar final velocity,
although we find somewhat higher velocities due to our inclusion of
magnetic forces.

\subsection{Euler Integrations for the Clouds}
We are now in a position to calculate the motion of the clouds as they
are pushed along by a mix of radiation pressure and the drag force
felt through their relative motion with respect to the wind (the
clouds are assumed to be diamagnetic, and thus are only confined by
the magnetic field; they are not guided or accelerated by the magnetic
fields in the continuous wind).  The effect of the drag force is very
important in two different ways.  First, it is the primary source of
cloud acceleration near the disk where the continuous wind launches
the cloud by drag forces.  In this case, the wind pulls the clouds
along via drag, which slows the wind down.  However, when the clouds
feel significant radiative acceleration, the drag forces change sign,
and the clouds help push the wind.

\subsubsection{Equation of Motion}

        We integrate the following equations of motion for the clouds
\begin{eqnarray}
\dot{v}_{r,cloud} & = & \frac{C_F \rho_{wind} R_{cloud}^2}{M_{cloud}} |
\mathbf{v}_{wind} - \mathbf{v}_{cloud}| (v_{wind,r} - v_{cloud,r}) - [1 -
\Gamma(\theta)] \frac{G M_{\bullet}}{r^2 + z^2} sin~\theta \\
\dot{v}_{\phi, cloud} & = & \frac{C_F \rho_{wind} R_{cloud}^2}{M_{cloud}} |
\mathbf{v}_{wind} - \mathbf{v}_{cloud}| (v_{wind,\phi} - v_{cloud,\phi}) \\
\dot{v}_{z, cloud} & = & \frac{C_F \rho_{wind} R_{cloud}^2}{M_{cloud}} |
\mathbf{v}_{wind} - \mathbf{v}_{cloud}| (v_{wind,z} - v_{cloud,z}) - [1 -
\Gamma(\theta)] \frac{G M_{\bullet}}{r^2 + z^2} cos~\theta,
\end{eqnarray}
where $R_{\rm cloud}$ is the radius of the cloud, $M_{\rm cloud}$ is
the mass of the cloud, and $C_F$ is the constant drag coefficient
\citep[for the solutions presented here, we use $C_F = 10$,][]{BK79},
and the overdot indicates derivatives with respect to time.  The only
non-trivial calculation is again the determination of the radiative
acceleration.

\subsubsection{Radiative Acceleration}

For the calculation of radiative acceleration on the clouds, we can
use the same equations that we used for the continuous wind, with two
exceptions: the definition of $t$ and the equation for $\frac{d
v_R}{dR}$.  

To calculate the effective electron optical depth ($t$) for the
clouds, we must take into account that the clouds do not fill all of
the wind.  We therefore define:
\begin{eqnarray}
t_{\rm cloud} \equiv \frac{\sigma_T \epsilon n_e v_{th}}{\frac{dv_R}{dR}},\label{tCloudEqn}
\end{eqnarray}
where $\epsilon$ is the fraction of the wind volume taken up by the
clouds \citep[see][]{Arav94}.  This factor is very important as the
matter in the clouds only occupies a small fraction of the wind
volume, and so the opacity is much smaller than it would be for the
continuous wind, for instance.

Once again, we must calculate $\frac{d v_R}{dR}$, where R is the
spherical radial coordinate.  We can go through a more rigorous
calculation than before by first recognizing that, since we are
following the clouds as a function of time, the total derivative is
\begin{eqnarray}
\frac{d v_R}{dR} & = & \frac{\partial v_R}{\partial t} \frac{\partial
t}{\partial R} + \frac{\partial v_R}{\partial r_0} \frac{\partial
r_0}{\partial R},
\end{eqnarray}
where $r_0$ is the cylindrical radius of the footpoint where the cloud
started to rise from the disk.  Since $v_R = v_p \cos (\theta -
\theta_F)$, we write
\begin{eqnarray}
\frac{\partial v_R}{\partial t} & = & \frac{1}{v_p} \left( v_r \frac{d
v_r}{dt} + v_z \frac{d v_z}{dt} \right) \cos(\theta - \theta_F) +
\nonumber \\ & & - v_p \sin(\theta - \theta_F) \left( \frac{d\theta}{dt}
- \frac{d\theta_F}{dt} \right).
\end{eqnarray}
The derivatives of angles $\theta$ and $\theta_F$ are
self-consistently computed from their definitions
\begin{eqnarray}
\theta & = & \tan^{-1}\left( \frac{r}{z} \right) \\
\Rightarrow \frac{d\theta}{dt} & = & \frac{1}{1 + (r/z)^2} \left(
\frac{1}{z} \frac{dr}{dt} - \frac{r}{z^2} \frac{dz}{dt} \right) 
= \frac{1}{1 + (r/z)^2} \left(
\frac{1}{z} v_r - \frac{r}{z^2} v_z \right),
\end{eqnarray}
and
\begin{eqnarray}
\theta_F & = & \tan^{-1}\left( \frac{v_r}{v_z} \right) \\
\Rightarrow \frac{d\theta_F}{dt} & = & \frac{1}{1 + (v_r/v_z)^2} \left(
\frac{1}{v_z} \frac{d v_r}{dt} - \frac{v_r}{v_z^2} \frac{d v_z}{dt}
\right). 
\end{eqnarray}

Meanwhile, for the other terms in the equation, we write
\begin{eqnarray}
\frac{\partial t}{\partial R} & = & \frac{1}{v_R} \\
\frac{\partial v_R}{\partial r_0} & = & -\frac{v_R}{2 r_0},
\end{eqnarray}
where, for the last equation, we assume that the gradient of the
velocity with respect to the starting radius scales according to the
Keplerian relation.  

Finally, to calculate the derivative $\frac{\partial r_0}{\partial
R}$, we assume that the flowlines do not change direction rapidly
along the spherical, radial line of sight.  Then, from trigonometry,
we find
\begin{eqnarray}
\frac{\partial r_0}{\partial R} & = & \sin \theta - \cos \theta
\tan \theta_F 
\end{eqnarray}
Bringing all of those terms together, and simplifying, we have the
following expression:
\begin{eqnarray}
\frac{d v_R}{dR} & = & \frac{v_r \dot{v}_r + v_z \dot{v}_z}{v_p^2} -
 \frac{\sin(\theta - \theta_F)}{\cos(\theta - \theta_F)} \left(\dot{\theta}
- \dot{\theta}_F \right)  - \frac{v_p \cos (\theta - \theta_F)}{2 r_0}
\left(\sin \theta - \cos \theta
\tan \theta_F \right).
\end{eqnarray}

\subsubsection{Calculating Drag Forces on the Wind}

We next integrate the cloud equations of motion, recording the
position and velocity of the clouds as a function of angle above the
disk.  We also output the radiative force on the clouds and, in order
to calculate the drag on the continuous wind, the drag forces of the
clouds on the wind.  We calculate those drag forces almost exactly the
same way we computed the drag force of the wind on the clouds, except
that now we must include the effect of many clouds on the wind, taking
into account the ensemble density of clouds.

The radial component of the wind's drag force on the clouds is given
by the following force equation (for simplicity, we follow just the
radial equation, here; the other components [azimuthal and vertical]
are very similar except for the final term which is the difference of
that component's velocities):
\begin{eqnarray}
F_{r,drag,cloud} & = & C_F \rho_{wind} R_{cloud}^2
|\mathbf{v}_{wind} - \mathbf{v}_{cloud}| (v_{wind,r} - v_{cloud,r}).
\end{eqnarray}
The force of any given cloud back on the wind is equal and opposite to
this,
\begin{eqnarray}
F_{r,drag,wind} & = & - C_F \rho_{wind}
R_{cloud}^2 |\mathbf{v}_{wind} - \mathbf{v}_{cloud}| (v_{wind,r}
- v_{cloud,r}).
\end{eqnarray}
To find the acceleration from this drag force, which we use as an
additional acceleration in our self-similar continuous wind model, we
then write
\begin{eqnarray}
\rho_{wind} V_{wind~per~cloud} a_{r,wind} & = & - C_F \rho_{wind}
R_{cloud}^2 |\mathbf{v}_{wind} - \mathbf{v}_{cloud}| (v_{wind,r}
- v_{cloud,r}),
\end{eqnarray}
where $V_{wind~per~cloud}$ is the wind volume per cloud,
$V_{wind~per~cloud} = n_{cloud,ens}^{-1}$.  Substituting this in
yields: 
\begin{eqnarray}
\rho_{wind} n_{cloud,ens}^{-1} a_{r,wind} & = & - C_F \rho_{wind}
R_{cloud}^2 |\mathbf{v}_{wind} - \mathbf{v}_{cloud}| (v_{wind,r}
- v_{cloud,r}).
\end{eqnarray}
After some simplification, 
\begin{eqnarray}
a_{r,wind} & = & - C_F n_{cloud,ens}
R_{cloud}^2 |\mathbf{v}_{wind} - \mathbf{v}_{cloud}| (v_{wind,r}
- v_{cloud,r}) 
\end{eqnarray}
For interface into the self-similar wind models, we just divide by the
gravitational acceleration, so this force can be included in the same
way that $\Gamma(\theta)$ is included in the self-similar model.
Therefore, we define
\begin{eqnarray}
\Delta_{r} & = & \frac{- C_F n_{cloud,ens}
R_{cloud}^2 |\mathbf{v}_{wind} - \mathbf{v}_{cloud}| (v_{wind,r}
- v_{cloud,r})}{G M_{\bullet}/(r^2 + z^2)}, 
\end{eqnarray}
where the other components of the drag equation would yield similar
expressions for $\Delta_{\phi}$ and $\Delta_{z}$, the drag in the
azimuthal and vertical directions.

The function $\Delta_r(\theta)$ is input back into both the
self-similar MHD wind solver as well as the program that computes the
continuous wind's radiative acceleration.  In the case of the
self-similar wind, we must be careful to only include the spherical
radial component of the drag force, in order to not disturb the
self-similar nature of those equations (since we are including this
force as a modification to gravity, it must also scale as gravity,
depending only on the spherical radius $R$).  Thus, the quantity input
back into the self-similar wind model is:
\begin{eqnarray}
\Delta_{R} & = & \Delta_r \sin \theta  + \Delta_z \cos \theta.
\end{eqnarray}

In the program that integrates Euler's equation to find the continuous
wind's radiative acceleration, we also include the drag force, but in
a slightly different way.  Since that program computes the poloidal
velocity of the wind, $v_p$, it uses only the poloidal component of
the drag force in modifying its force laws.  Within that program, we
therefore calculate
\begin{eqnarray}
\Delta_{P} & = & \Delta_r \sin \theta_F + \Delta_z \cos \theta_F.
\end{eqnarray}

These accelerations are included in successive iterations of the code.



\section{Early Results}\label{Results}

In this section, we outline further tests of our model, and end by
using the model to explore the dependence of the wind on key input
parameters.  

\subsection{Tests of Code Convergence}

\begin{figure}
\begin{center}
\includegraphics[angle=-90, width=6in]{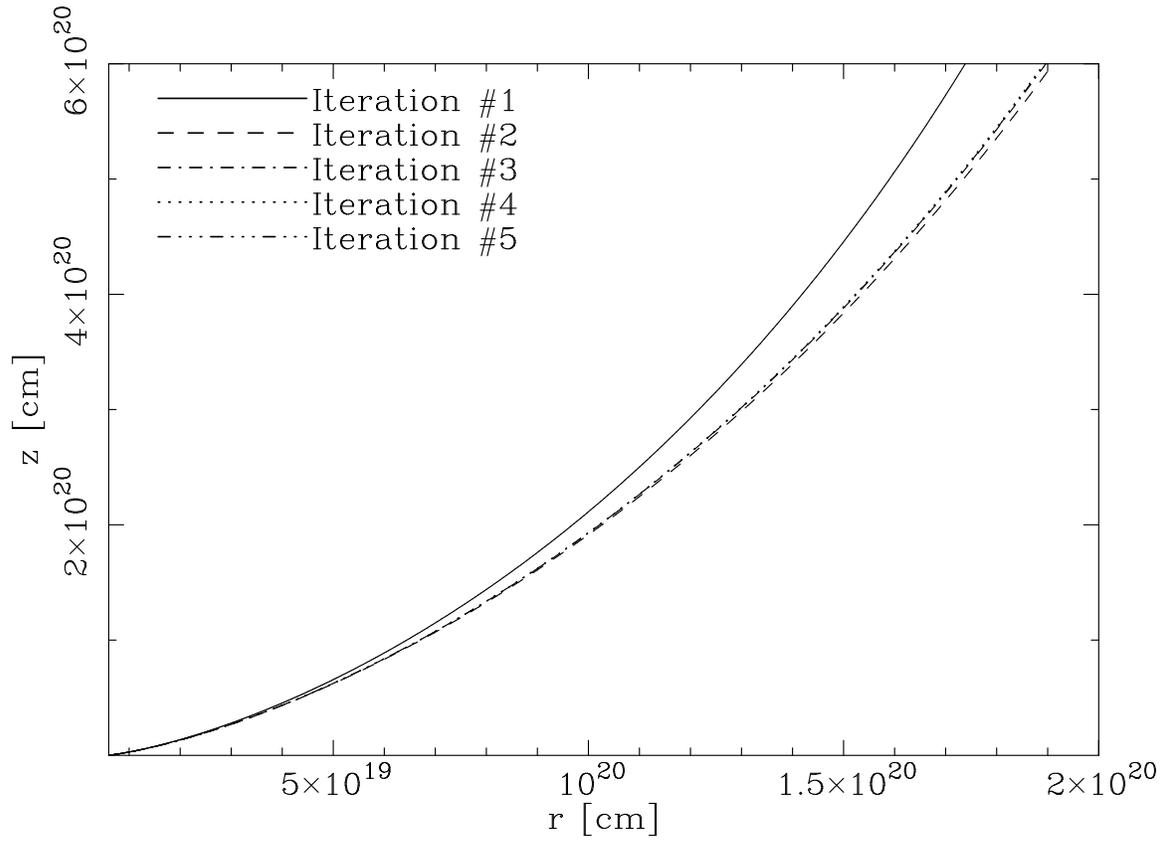}
\caption{A plot of the poloidal wind streamlines in our model and
their evolution over several iterations of the code.  After the third
iteration, the solution does not change, and the lines overlap,
indicating convergence for this outflow.\label{windIterations}}
\end{center}
\end{figure}

	We first show that with this model setup, the iterations in
the model converge to a stable solution, after which successive
iterations of the code do not significantly alter the structure of the
wind (see Fig.~\ref{windIterations}).  Almost as important, the system
converges to a solution within just a few iterations, corresponding to
approximately a day of computational time on a standard desktop
workstation.

\subsection{Testing Other Radiative Acceleration Approximations}\label{radTests}

\begin{figure}
\begin{center}
\includegraphics[angle=-90,width=6in]{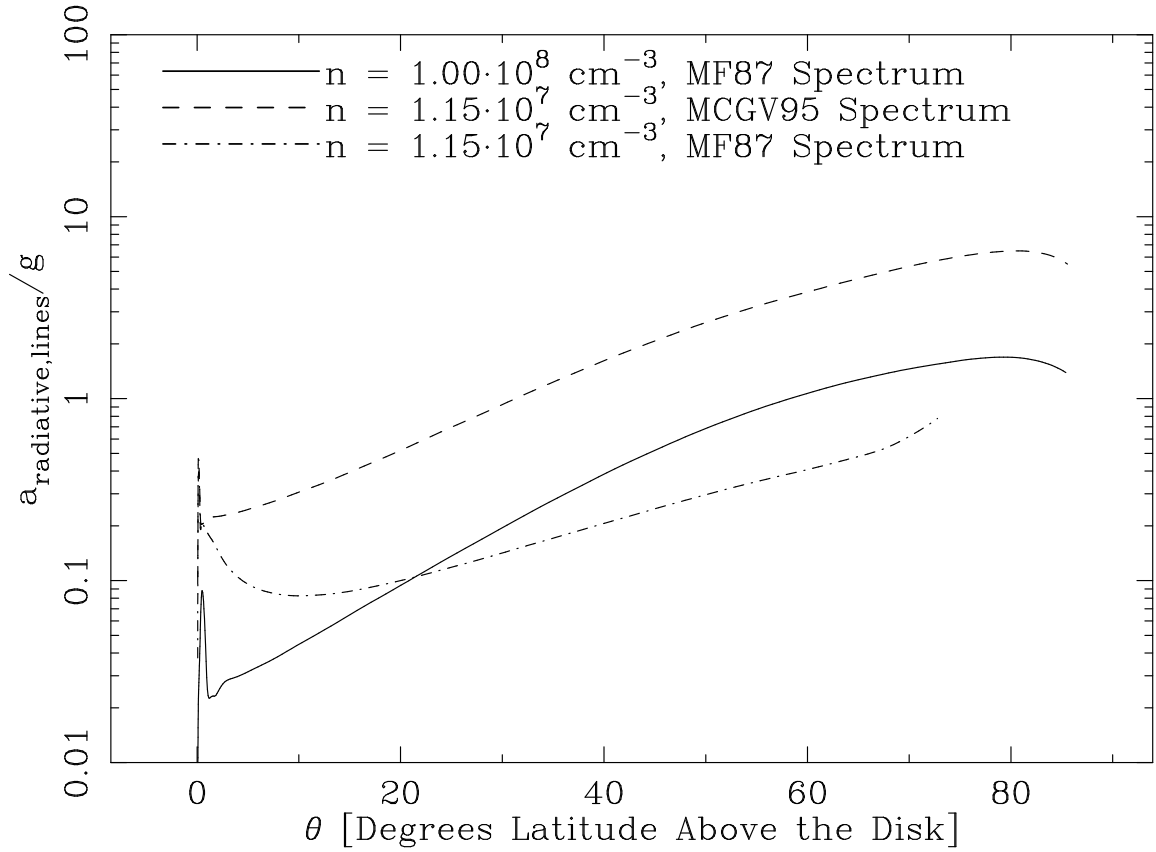}
\caption{A comparison of the radiative acceleration obtained with
different incident spectrum, showing the MF87 spectrum and the
spectrum assumed in MCGV95. For the initial wind density of
$10^7$~cm$^{-3}$, the MCGV95 spectrum yields much greater acceleration
than the Mathews \& Ferland spectrum.  Note that this solution does
not include clouds, allowing a better comparison with the continuous
wind simulations of MCGV95.
\label{incidentSpectrumCompare}}
\end{center}
\end{figure}

	Our early runs with this model have given interesting results
when comparing this work to previous research on AGN outflows.  One of
our first checks on the radiative acceleration code was to compare
with previous published results to see if we obtain similar
acceleration factors.  We found that, considering the continuous
wind's radiative acceleration alone and assuming an MF87 input
spectrum, the wind experienced much less acceleration than with the
continuum from MCGV95 (see Fig.~\ref{incidentSpectrumCompare}).  This
agrees well with the results of MCGV95, and shows the sensitivity of
radiative acceleration not only to the incident spectrum, but also to
the shielding that lies between the outflowing wind and the central
source, which has a strong impact on the spectrum incident on the
outflowing gas.

\begin{figure}[ht]
\begin{center}
\includegraphics[angle=-90,width=6in]{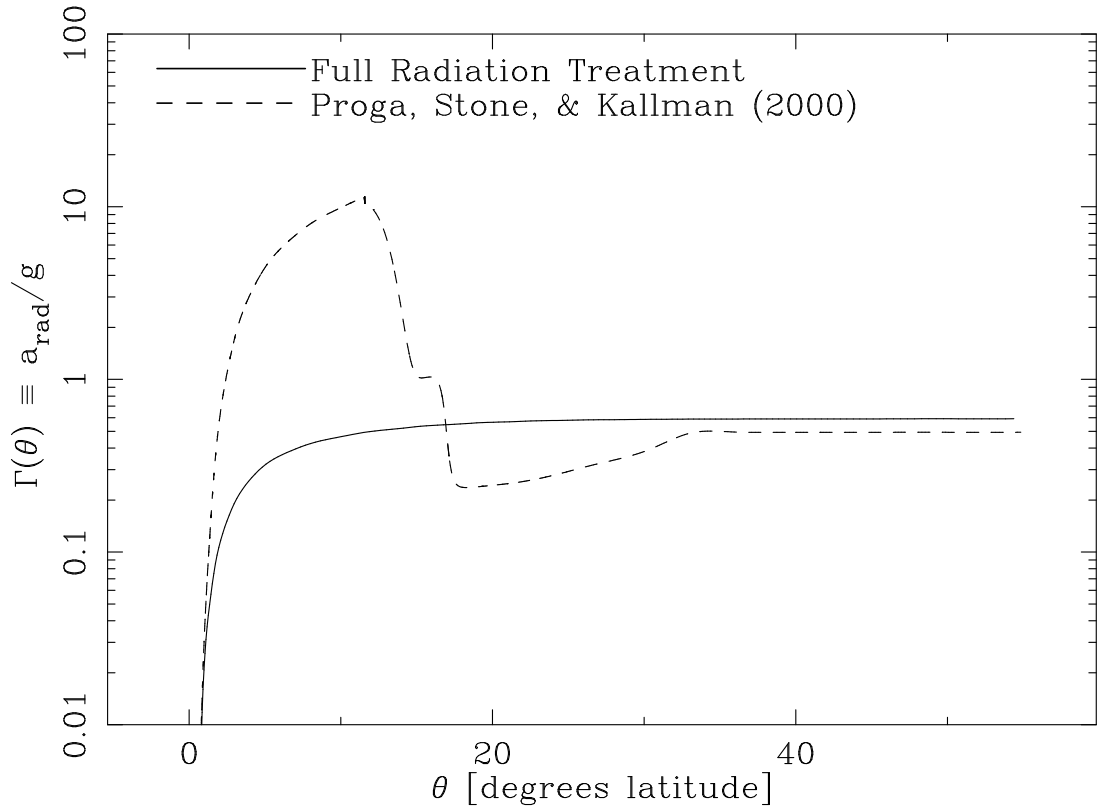}
\caption{A comparison between our full radiation treatment and
calculations equivalent to those used given in
\citet{Pro00}.\label{progaRadCompare}}
\end{center}
\end{figure}

	In another comparison with previous research, we re-programmed
the model to calculate radiative acceleration using the same
approximations as were used in \citet{Pro00}.  We were also careful to
use the same incident 10 keV bremsstrahlung spectrum that was assumed
in their work when they employed the results of \citet{SK90}. When we
inserted that radiation handling into our model, we found very
different radiative acceleration profiles than shown in our
calculations using Cloudy photoionization results.  This difference
could be due to a couple of factors.  First, in \citet{Pro00}, it is
also assumed that the fraction of flux in the X-ray and UV portions of
the spectrum were equal (both set to half of the incident flux).  As
was shown later \citep{Pro02a}, the balance of X-ray and UV radiation
in the incident spectrum is critically important: while the UV
radiation is the primary source of momentum for the gas, the X-ray
radiation ionizes the gas, and the more X-ray radiation present in the
spectrum, the fewer lines remain in the gas to absorb the momentum
from UV photons.  If the 10 keV bremsstrahlung spectrum does not have
this balance of UV and X-ray flux, it could be a source of error in
the calculations.  Second, the 2D hydro simulations only use estimates
of the ionization parameter rather than calculate it exactly.
	
	We must point out here that our goal is to test different
methods of computing radiative acceleration and check whether
approximations give the same result as our {\it ab initio}
calculations.  From this study, within the framework and assumptions
of our model, we can only conclude that the approximations used in
\citet{Pro00} do not match the radiative acceleration calculated from
first principles.  However, it is also important to realize that such
approximations allow those researchers the ability to simulate
larger-scale flows, which can lead to valuable insight.  We believe,
though, that these figures show the necessity of accurately handling
the radiation transfer and acceleration calculations.

In the subsection that follows, we perform a parameter survey, noting
the cases where radiative acceleration significantly affects the
geometry and kinematics of our wind models.

\begin{deluxetable}{lll}
\small 
\tablenum{1} 
\tablecaption{Fiducial parameters adopted for the
models in this paper.\tablenotemark{a}
\label{paramTable}} 

\tablehead{
\colhead{Parameter} & \colhead{Fiducial Value} & \colhead{Parameter
Description} } 

\startdata 
$\kappa$ & 0.03 & dimensionless ratio of mass flux to\\ 
& & \hspace{0.25in}magnetic flux in the wind\\
$\lambda$ & 30.0& normalized total specific angular \\ 
& & \hspace{0.25in}momentum of the wind\\
$b$ & 1.5 & power-law describing variation of \\
& & \hspace{0.25in}density with spherical radius at \\
& & \hspace{0.25in}the base of the wind: $n \propto R^{-b}$ \\
$M_{\bullet}$ & $10^8 M_{\sun}$ & mass of the central black hole\\
$L_{\rm continuum}$ & $0.25~L_{\rm Edd}$ & luminosity of the central
continuum\\
Incident Spectrum & Mathews \& Ferland (1987) & Spectrum for the
central continuum \\
$r_0$ & $6 \times 10^{18}~{\rm cm}$ & launch radius of the continuous
wind \\
$n_0$ & $10^8~{\rm cm}^{-3}$& initial density of the continuous wind
\\
& & \hspace{0.25in}at the launch radius \\ 
$N_{\rm H,shield}$ & $10^{25}~{\rm cm}^{-2}$& gas shielding column at
the base of \\
& & \hspace{0.25in} the wind \\
Dust in Continuous Wind & No & presence of dust in the \\
& & \hspace{0.25in} continuous wind\\
$M_{\rm cloud}$ & $10^{-2} M_{\rm cloud, max}$ & mass of the cloud as
a fraction of the 
\\ & & \hspace{0.25in}maximum cloud mass the wind 
\\ & & \hspace{0.25in}can uplift by ram pressure \\
$\dot{M}_{\rm cloud}$ & $10^{-2} \dot{M}_{\rm wind}$ & mass outflow
rate of the clouds as a \\
& & \hspace{0.25in} fraction of the mass outflow \\
& & \hspace{0.25in} rate of the wind\\
$C_F$ & $10$ & dimensionless geometrical drag \\
& & \hspace{0.25in} coefficient of the clouds
\enddata
\tablenotetext{a}{Any changes to these parameters are listed in the
figure captions.}
\end{deluxetable}

\begin{deluxetable}{lll}
\small
\tablenum{2} 
\tablecaption{Range of values examined in the parameter
survey and relevant figures for each parameter\label{surveyTable}}
\tablehead{ 
\colhead{Parameter} & \colhead{Range of Values} & \colhead{Figure(s)}}
\startdata 
$M_{\rm cloud}$ & $ 0.01$ -- $0.1 M_{\rm cloud, max}$ & \ref{WindVaryCloud} \\
$\dot{M}_{\rm cloud}$ & $0.01$ -- $0.5 \dot{M}_{\rm wind}$ &
\ref{WindVaryCloud} \\
$r_0$ & $6 \times 10^{17}$ -- $6 \times 10^{19}~{\rm cm}$
& \ref{WindVaryR}, \ref{Windr6e17} \\
$n_0$ & $10^6$ -- $10^{10} {\rm cm}^{-3}$ & \ref{WindVaryDensity} \\
$N_{\rm H,shield}$ & $10^{22}$ -- $10^{26} {\rm cm}^{-2}$
&\ref{WindVaryShield1e9}, \ref{WindVaryShield1e8} \\
Dust in Continuous Wind & No Dust, ISM Dust, Orion Dust &\ref{WindVaryDust} \\
$L_{\rm continuum}$ & $0.1$ -- $0.5~L_{\rm Edd}$ &
\ref{WindVarySpectrum1e9}, \ref{WindVarySpectrum1e8} \\
Incident Spectrum & MF87, MCGV95 &
\ref{WindVarySpectrum1e9}, \ref{WindVarySpectrum1e8} \\
$\kappa$ & $0.015$ -- $0.1$ & \ref{WindVaryKappa1e9}, \ref{WindVaryKappa1e8}
\\
$M_{\bullet}$ & $10^7$ -- $10^9 M_{\sun}$ & \ref{WindVaryMBH}\\
\enddata
\end{deluxetable}

\clearpage
\begin{figure}[p]
\begin{center}
\epsscale{0.80}
\plotone{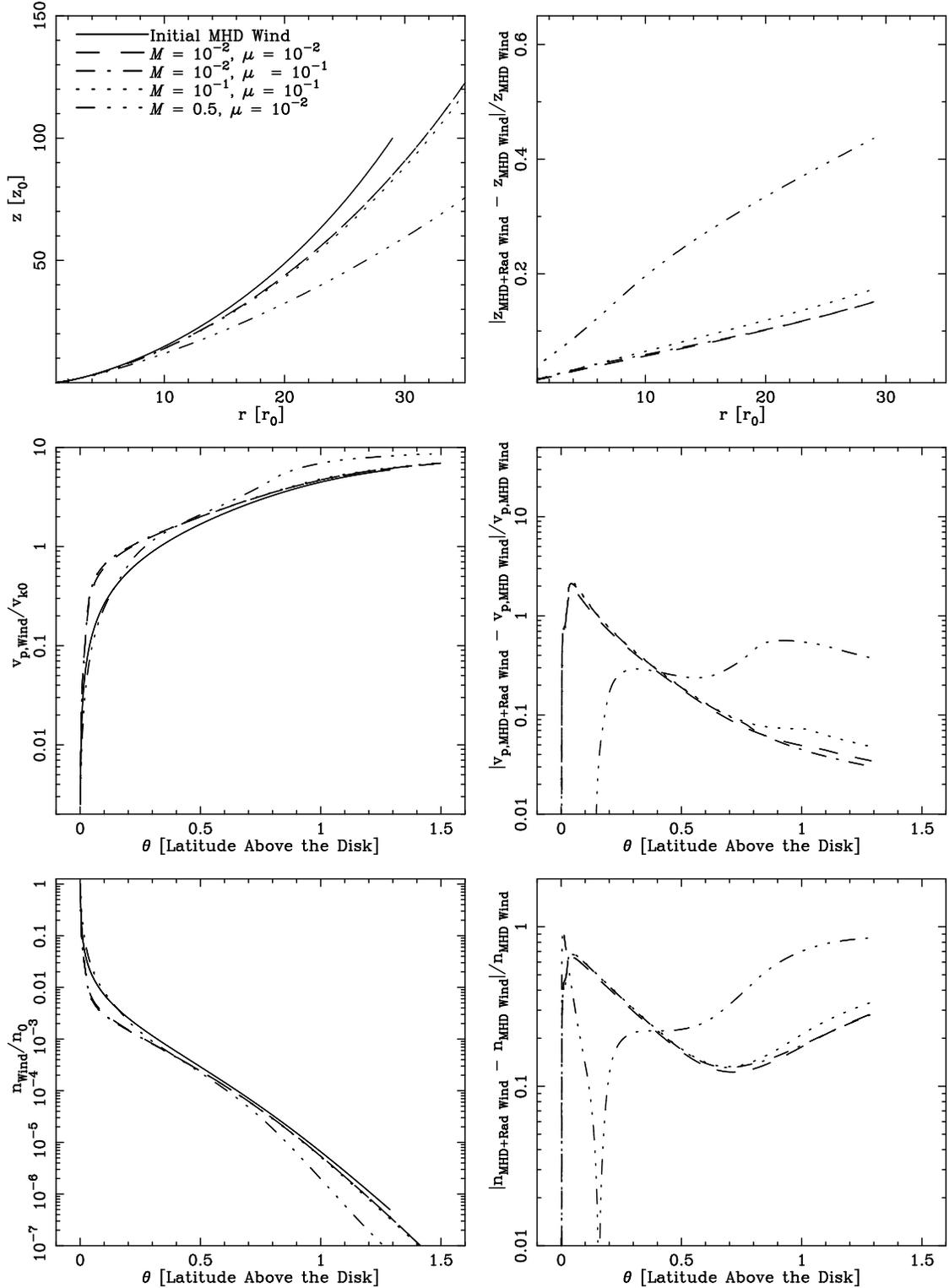}
\caption{The six panels of this figure display the variation in the
wind model with changes in the cloud parameters: we vary both the mass
outflow rate of the clouds and the mass of the individual clouds.  The
plots on the left give the poloidal streamlines of the wind, the
poloidal velocity, and the density of the wind.  For each of these
figures, the solid black line gives the initial MHD wind model
(without radiative acceleration) and the various broken lines give the
wind's equilibrium structure after radiative acceleration has been
applied.  In the three plots on the right, that equilibrium structure
is compared to the initial MHD wind structure by plotting the
difference between the initial and final wind model divided by the
initial MHD wind results.  For this plot, we define $M = \dot{M}_{\rm
cloud}/\dot{M}_{\rm wind}$ and the cloud mass $\mu = M_{\rm
cloud}/M_{\rm cloud,max}$.  The low initial velocity difference and
large drop in the variation of wind density in the high cloud outflow
case is due to the large drag exerted by the clouds on the wind when
the wind launches the clouds from the disk.
\label{WindVaryCloud}}
\end{center}
\end{figure}

\clearpage
\begin{figure}[p]
\begin{center}
\epsscale{0.80}
\plotone{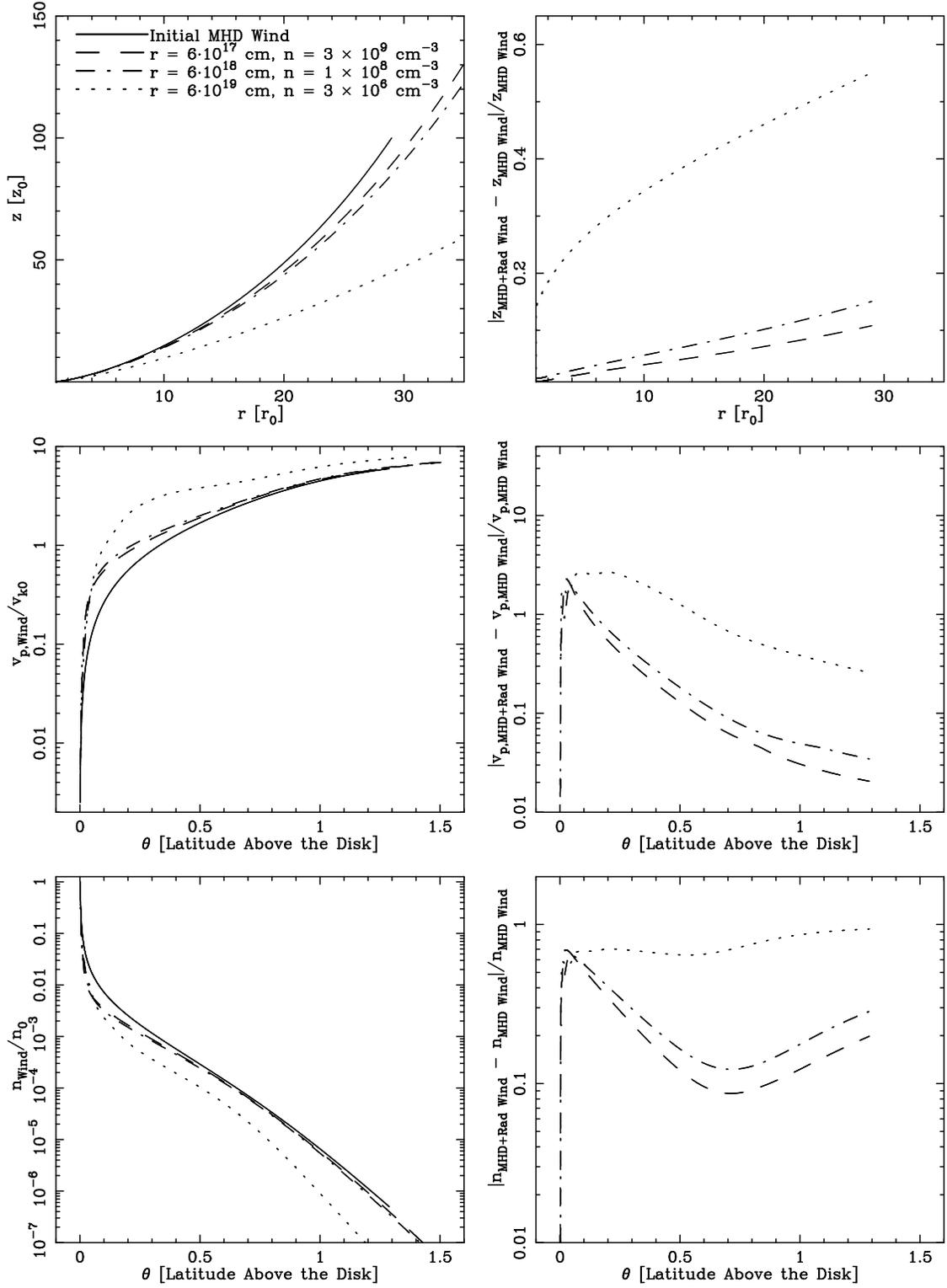}
\caption{As in Fig.~\ref{WindVaryCloud}, but showing the variation in wind
structure with changes in the launching radius of the continuous
wind.
\label{WindVaryR}}
\end{center}
\end{figure}

\clearpage
\begin{figure}[p]
\begin{center}
\epsscale{0.80}
\plotone{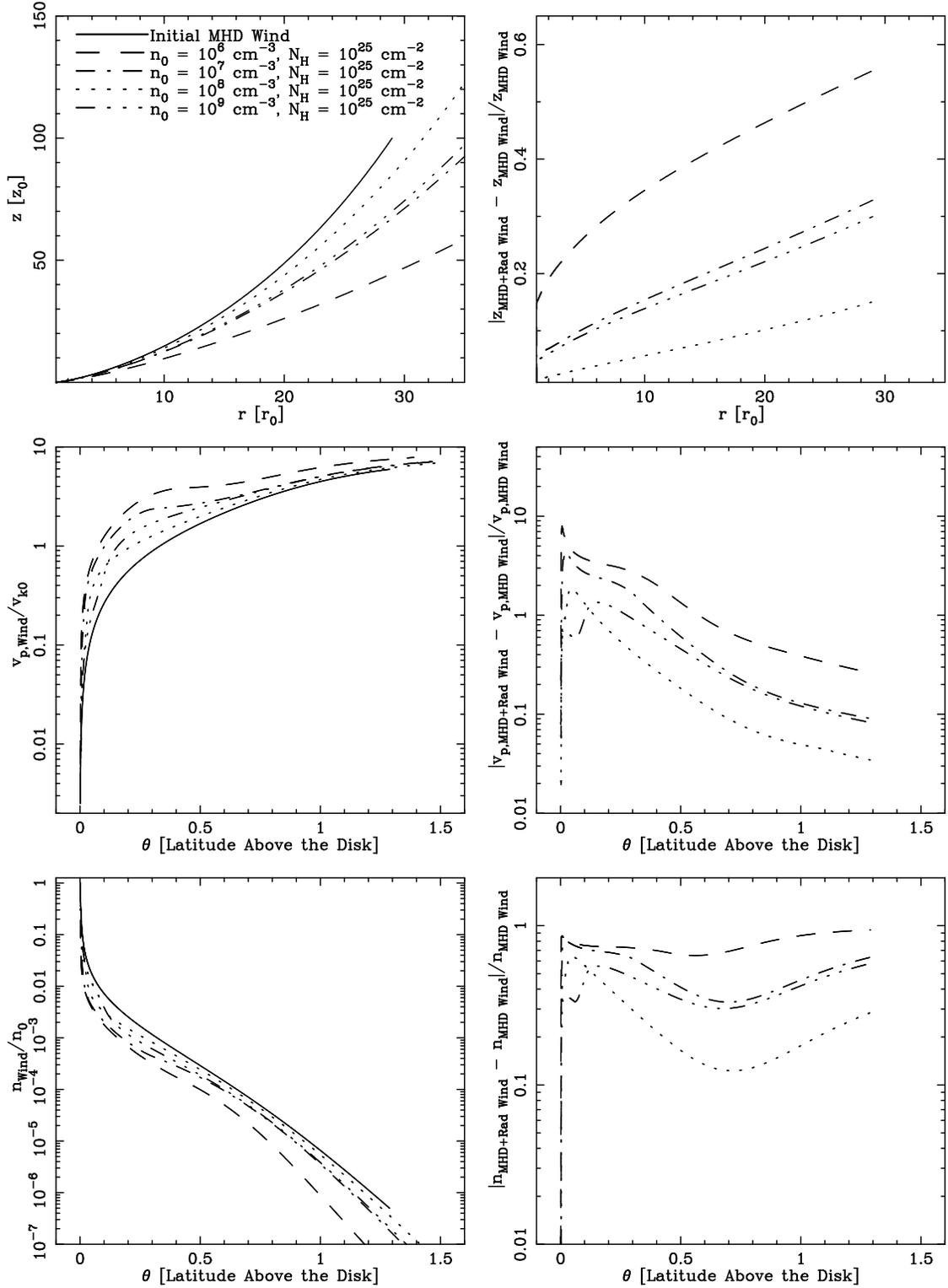}
\caption{As in Fig.~\ref{WindVaryCloud}, but showing the variation in
wind structure with changes in the initial density of the continuous
wind. 
\label{WindVaryDensity}}
\end{center}
\end{figure}

\clearpage
\begin{figure}[p]
\begin{center}
\epsscale{0.80}
\plotone{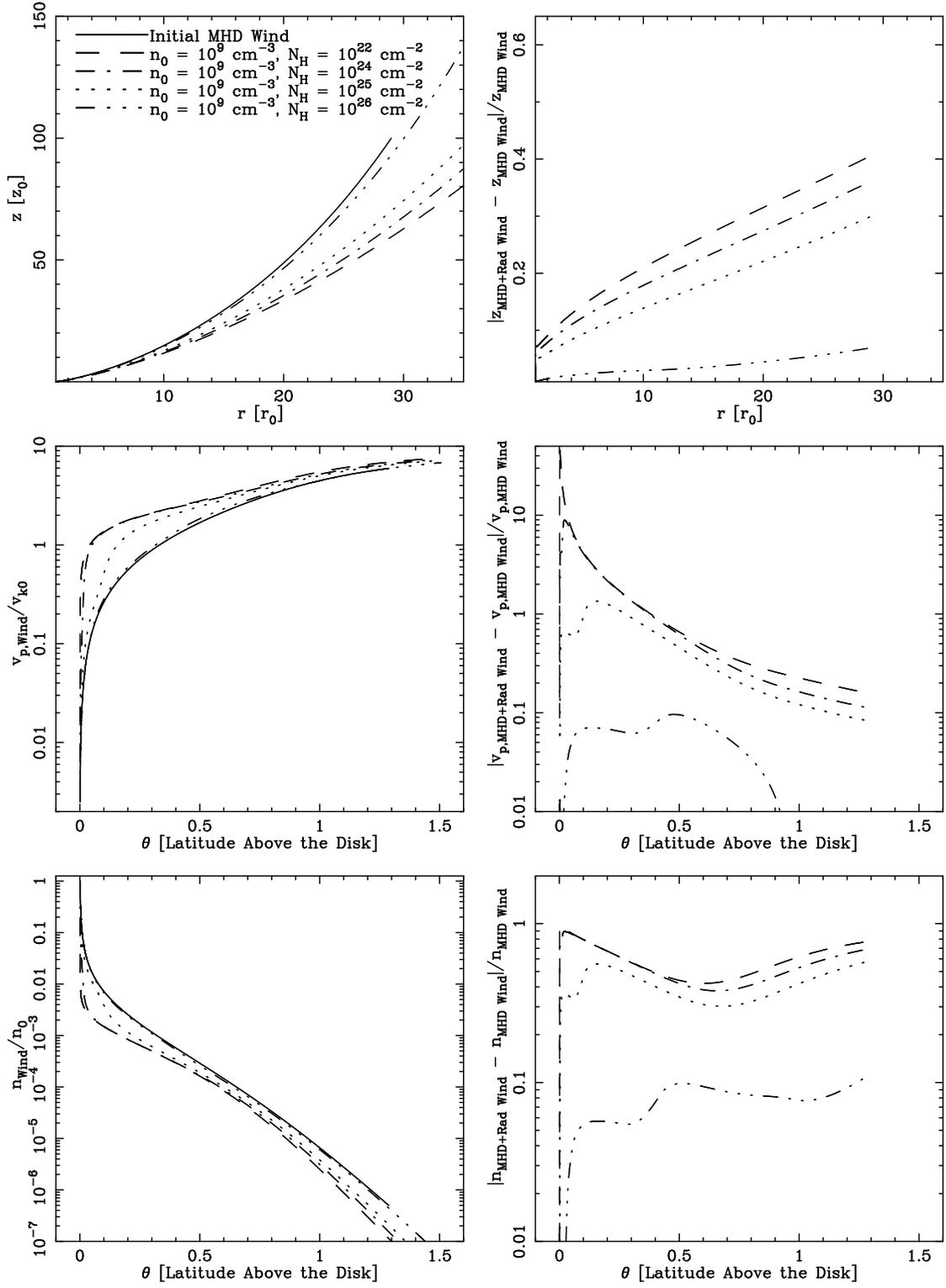}
\caption{As in Fig.~\ref{WindVaryCloud}, but showing the variation in
the wind structure with changes in the shielding gas column in front
of the wind, and with an initial wind density of $10^9$ cm$^{-3}$.
\label{WindVaryShield1e9}}
\end{center}
\end{figure}

\clearpage
\begin{figure}[p]
\begin{center}
\epsscale{0.80}
\plotone{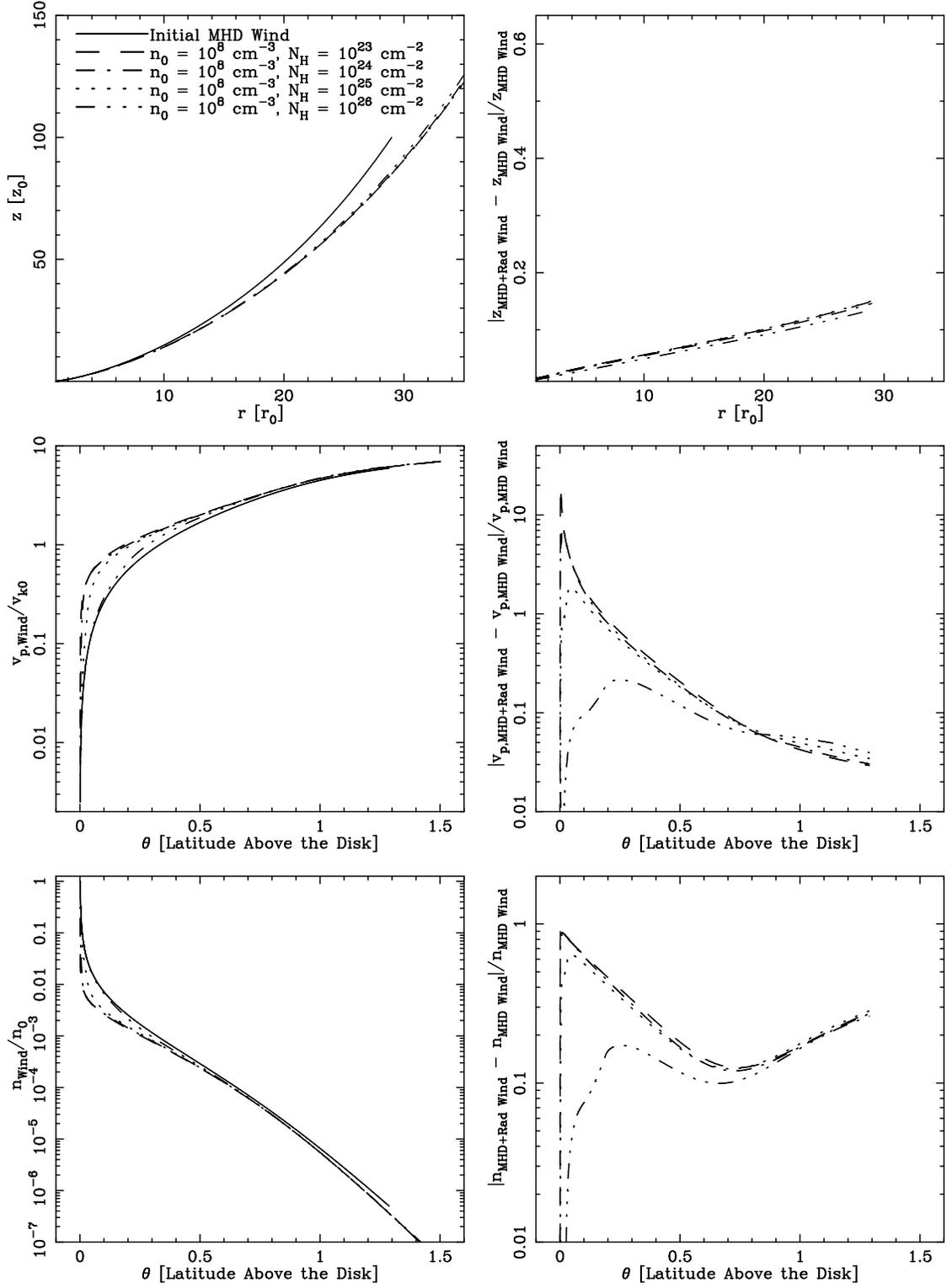}
\caption{As in Fig.~\ref{WindVaryCloud}, but showing different levels
of gas shielding in front of the outflow for an initial wind density
of $10^8$ cm$^{-3}$.  
\label{WindVaryShield1e8}}
\end{center}
\end{figure}

\clearpage
\begin{figure}[p]
\begin{center}
\epsscale{0.80}
\plotone{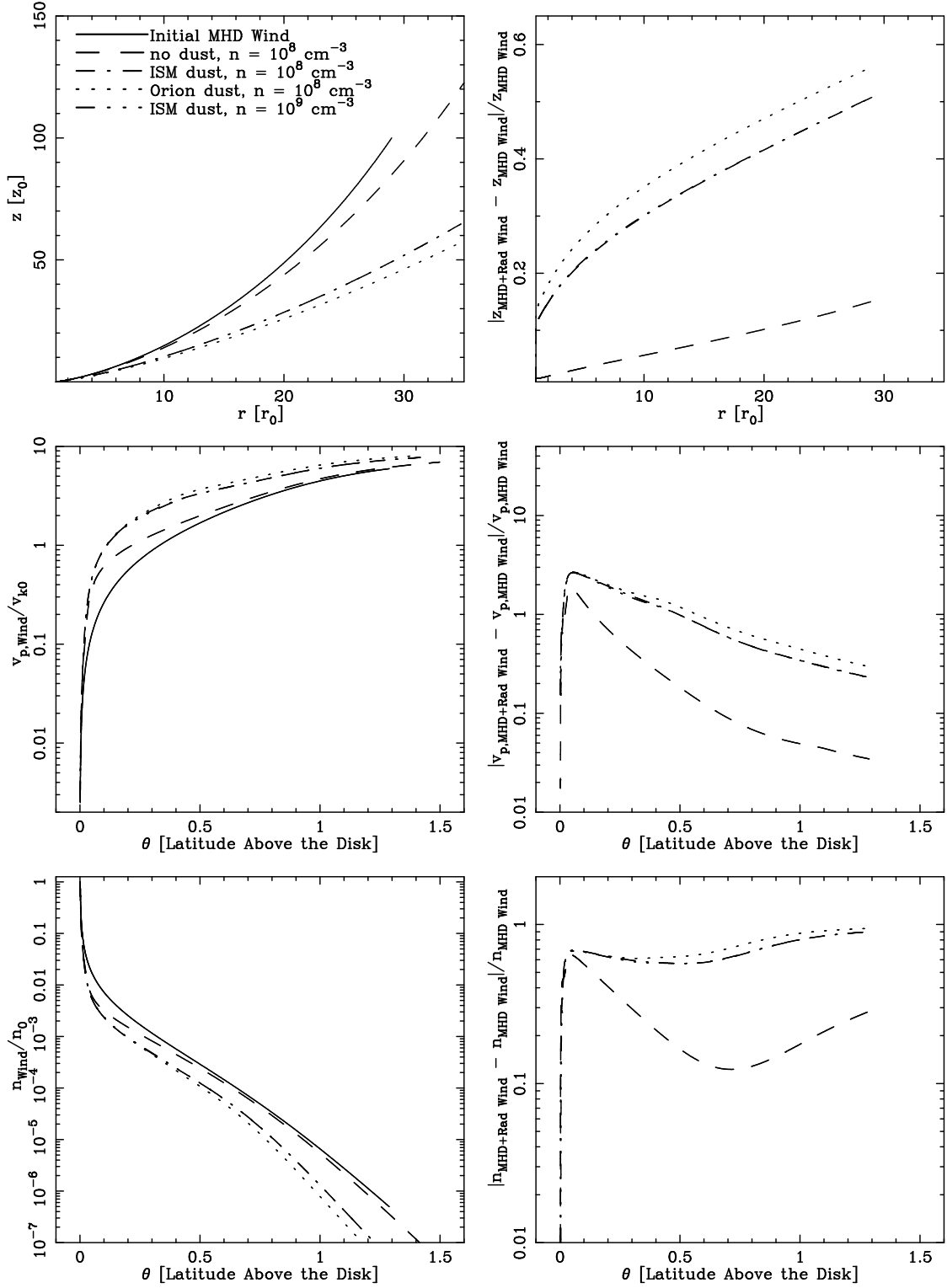}
\caption{As in Fig.~\ref{WindVaryCloud}, but showing the variation in
wind structure when dust is included in the continuous wind. 
\label{WindVaryDust}}
\end{center}
\end{figure}

\clearpage
\begin{figure}[p]
\begin{center}
\epsscale{0.80}
\plotone{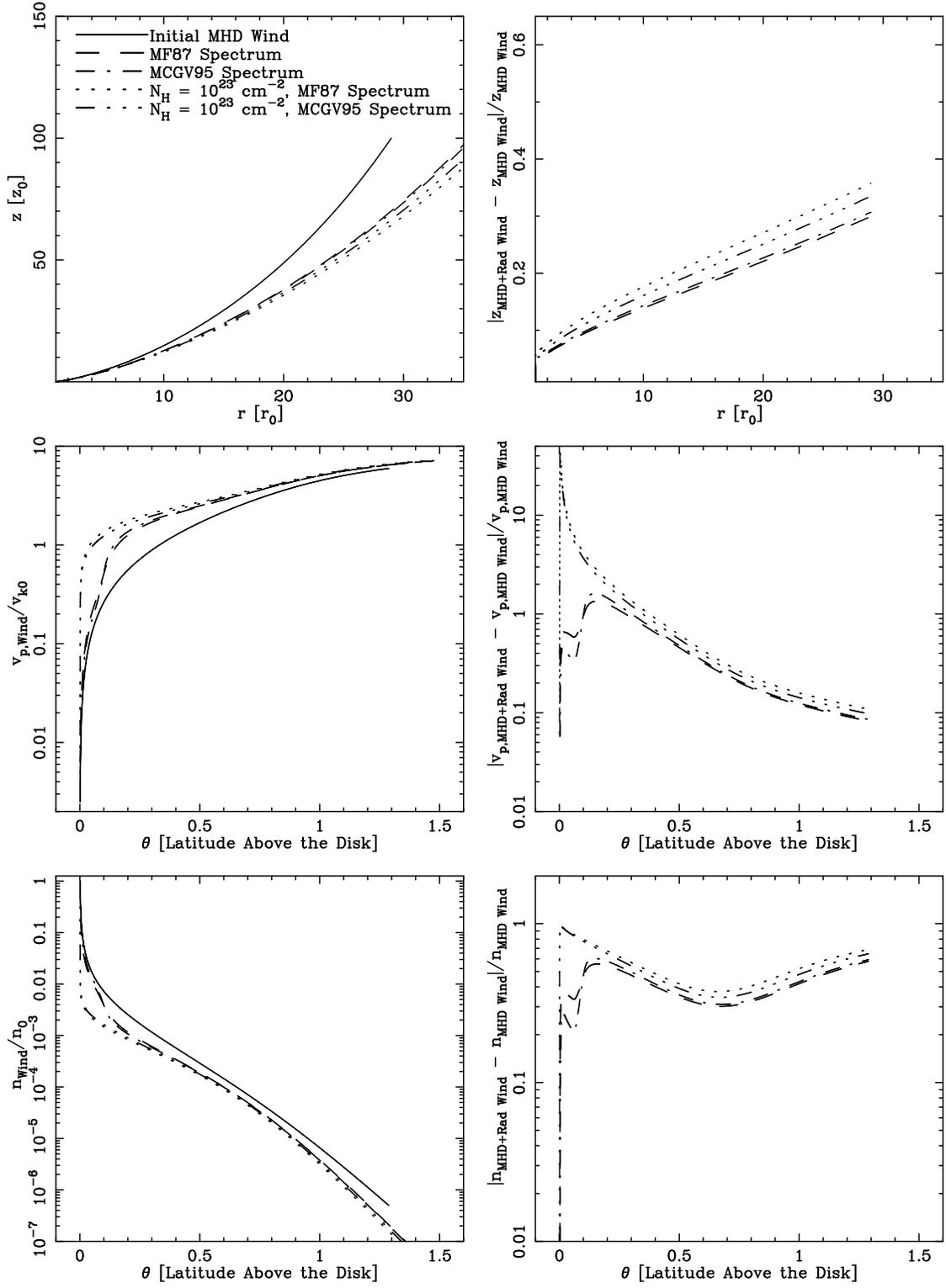}
\caption{As in Fig.~\ref{WindVaryCloud}, but showing the variation in
the wind structure with changes in the incident continuum and
shielding.  All of the solutions here were run with an initial wind
density of $n_0 = 10^9~{\rm cm}^{-3}$
\label{WindVarySpectrum1e9}}
\end{center}
\end{figure}

\clearpage
\begin{figure}[p]
\begin{center}
\epsscale{0.80}
\plotone{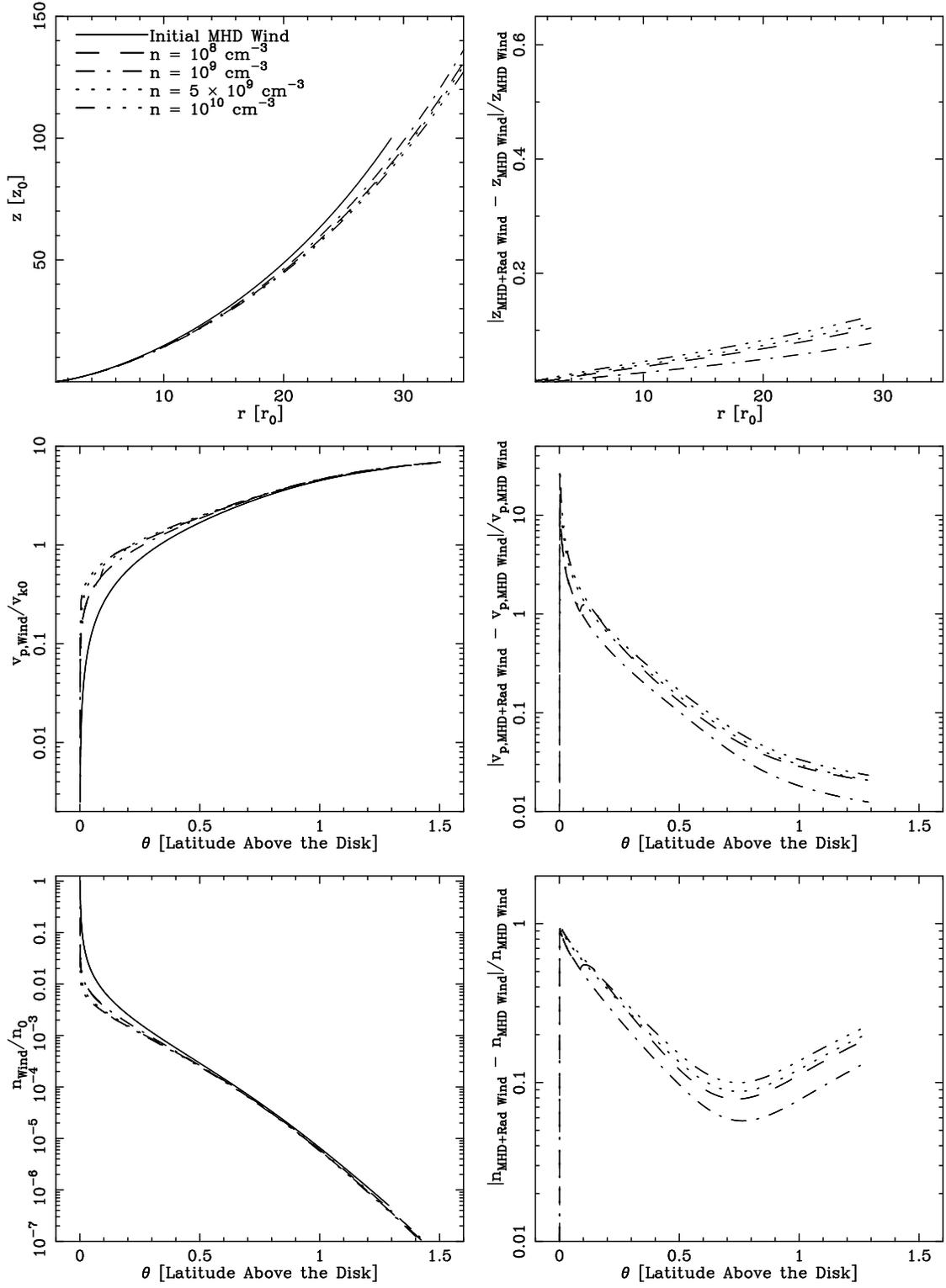}
\caption{As in Fig.~\ref{WindVaryCloud}, but showing the variation in
wind structure when the wind is launched at a radius an order of
magnitude closer to the central black hole.  All of the runs shown
here have an initial launch radius $r_0 = 6 \times 10^{17}~{\rm cm}$
and shielding columns of $N_H = 10^{23}~{\rm cm}^{-3}$.
\label{Windr6e17}}
\end{center}
\end{figure}

\clearpage
\begin{figure}[p]
\begin{center}
\epsscale{0.80}
\plotone{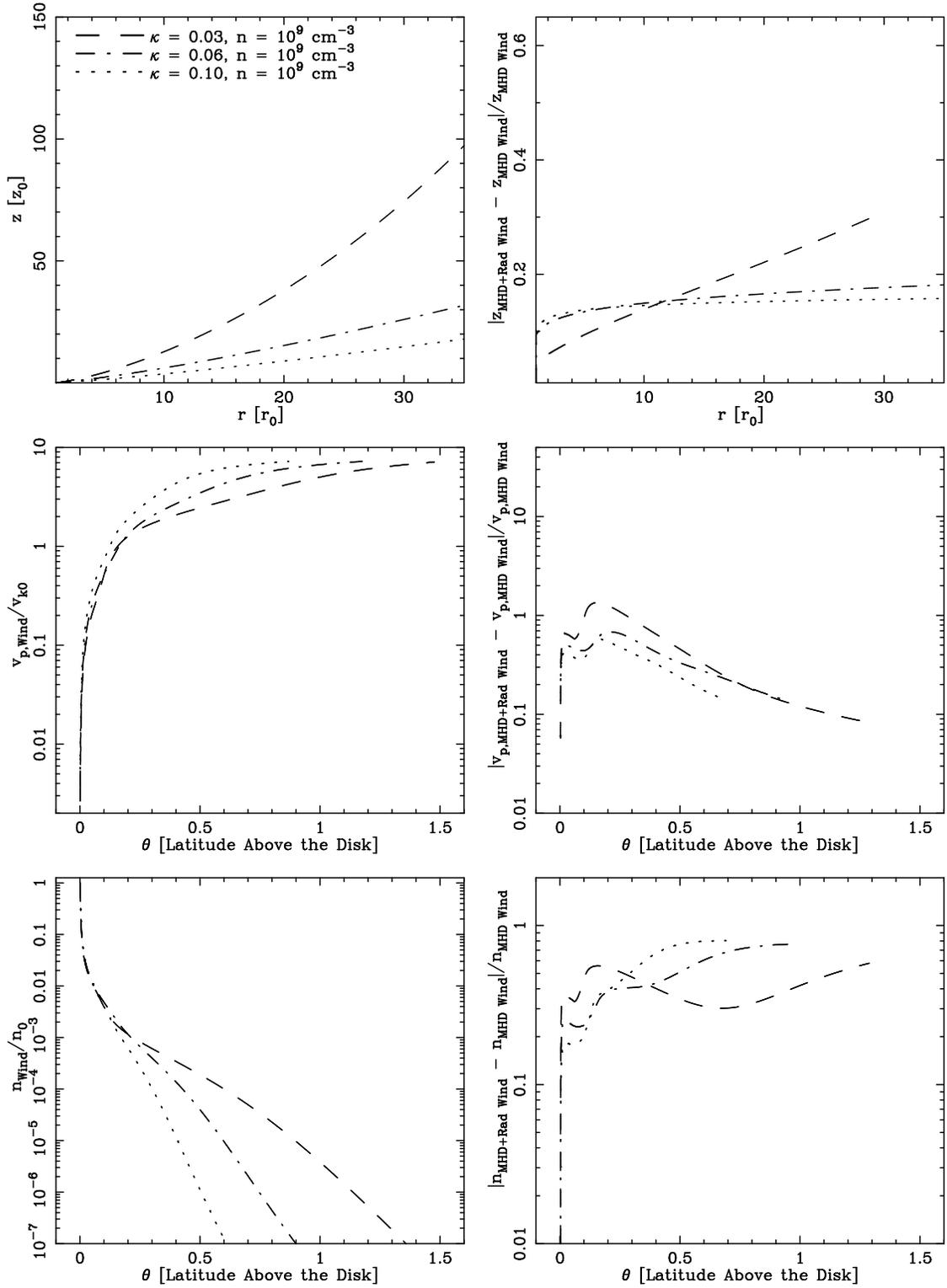}
\caption{As in Fig.~\ref{WindVaryCloud}, but showing the variation in
wind structure with changes in $\kappa$ (the dimensionless mass to
magnetic flux ratio in the wind) for wind models with an initial wind
density of $10^9$ cm$^{-3}$.  These plots differ from the other
parameter-variation plots given: as $\kappa$ is changed, the basic MHD
wind model changes, so each model shown has its own pre-radiative
acceleration MHD wind model.  To avoid confusion, therefore, none are
shown on the left panels, and each of the different plots on the right
compare the initial MHD wind model {\it for that value of $\kappa$}
and the final equilibrium model.  Also, note that the end-points of
the models (in altitude) are not the same for all models; this is
because the models are plotted as a function of $\theta$, and even
though the models all end at the same absolute height, $\theta$ is
smaller because the wind is less collimated and hence farther from the
central source, ending at a smaller value of $\theta$.
\label{WindVaryKappa1e9}}
\end{center}
\end{figure}

\clearpage
\begin{figure}[p]
\begin{center}
\epsscale{0.80}
\plotone{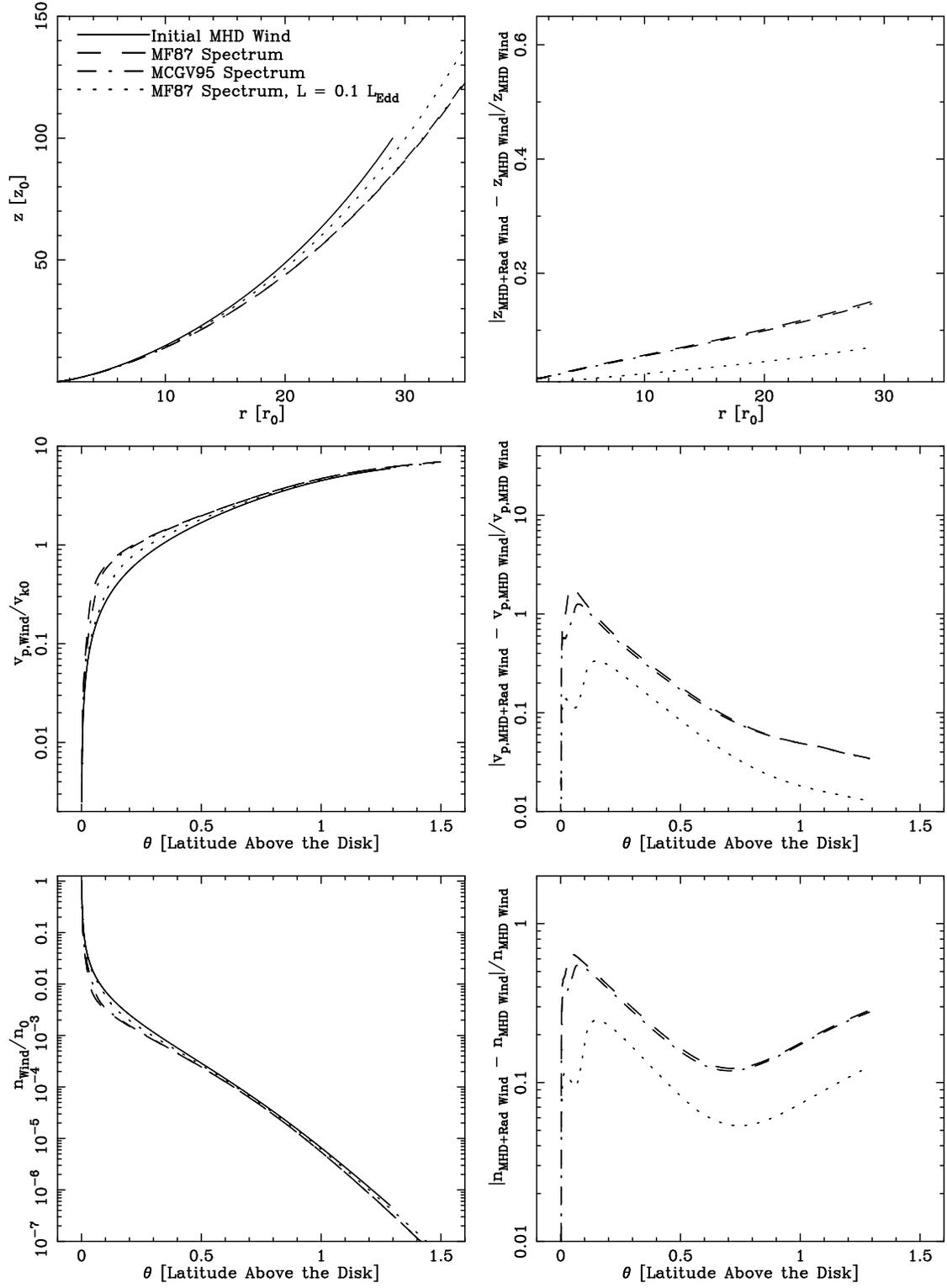}
\caption{As in Fig.~\ref{WindVaryCloud}, but showing the variation in
the wind structure with changes in the incident continuum.  All of the
solutions here have an initial wind density of $n_0 =
10^8~{\rm cm}^{-3}$
\label{WindVarySpectrum1e8}}
\end{center}
\end{figure}

\clearpage
\begin{figure}[p]
\begin{center}
\epsscale{0.80}
\plotone{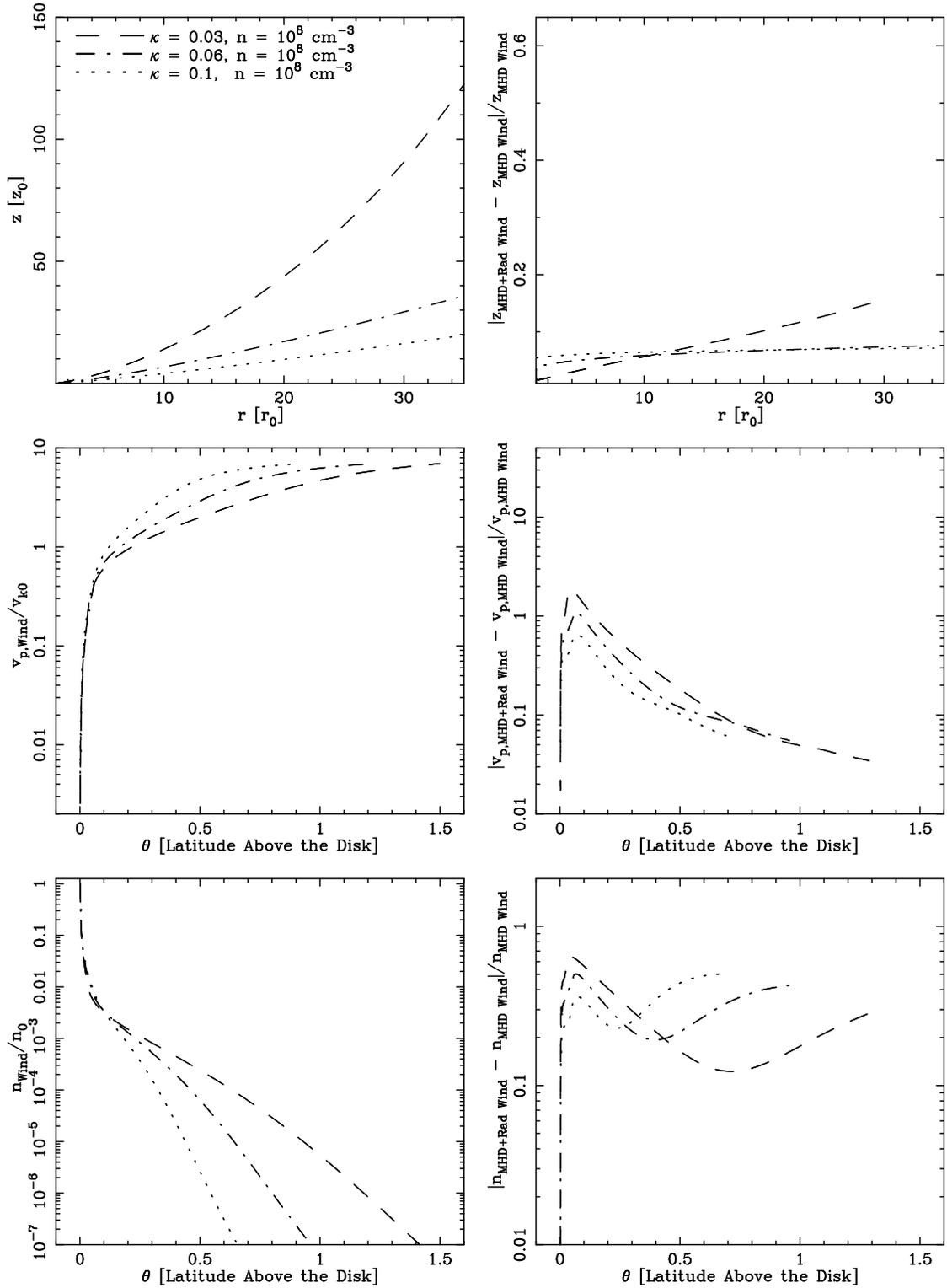}
\caption{As in Fig.~\ref{WindVaryKappa1e9}, but showing the variation
in wind structure with changes in $\kappa$ (the dimensionless mass to
magnetic flux ratio in the wind) for wind models with an initial wind
density of $10^8$ cm$^{-3}$.  \label{WindVaryKappa1e8}}
\end{center}
\end{figure}

\clearpage
\begin{figure}[p]
\begin{center}
\epsscale{0.80}
\plotone{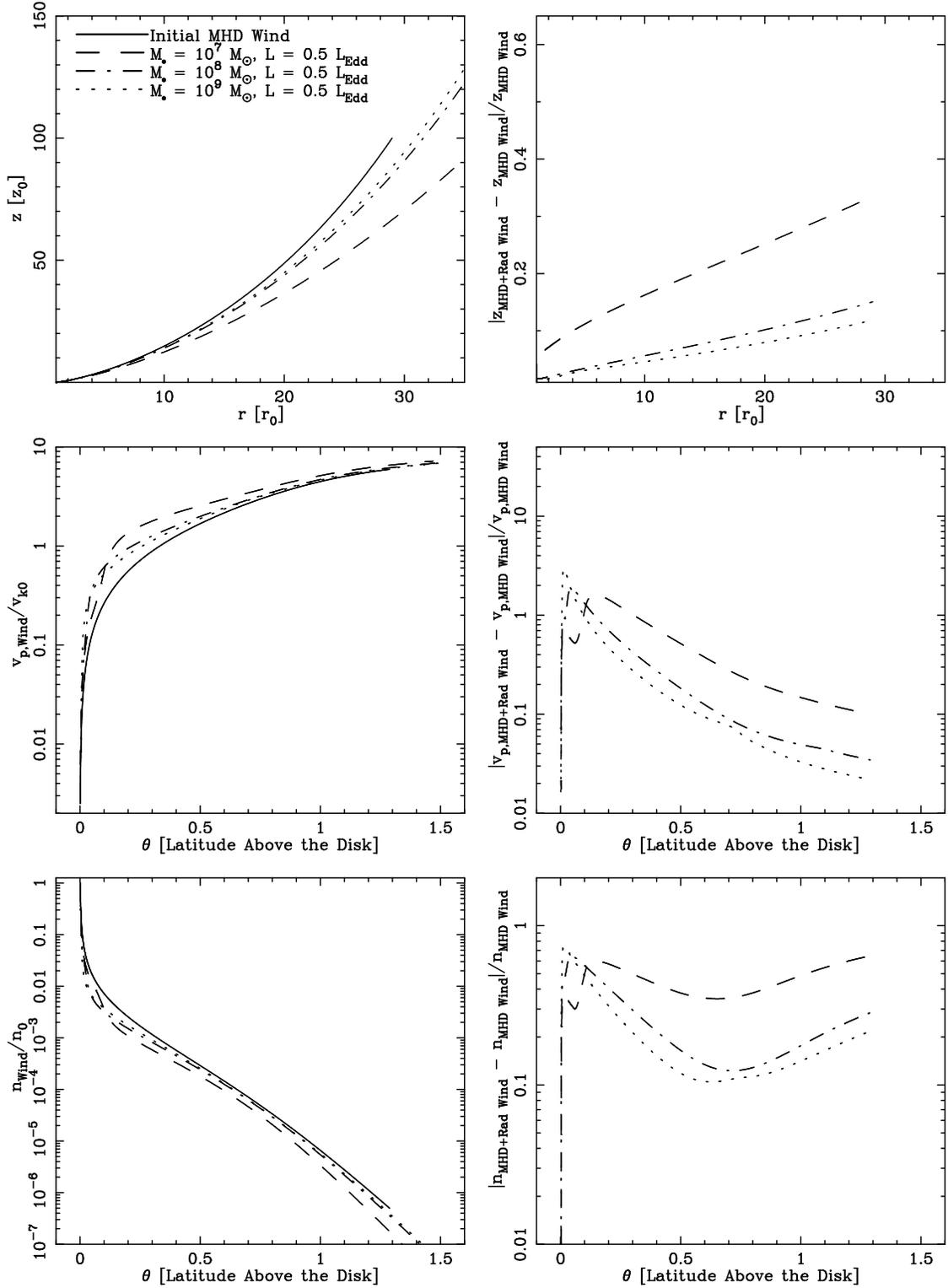}
\caption{As in Fig.~\ref{WindVaryCloud}, but showing the variation in
wind structure with changes in the mass of the central black hole.
\label{WindVaryMBH}}
\end{center}
\end{figure}

\subsection{Parameter Survey}\label{paramSurvey}

\subsubsection{Setup}

In this parameter survey, we compare the variations in the wind's
velocity, density, and radiative acceleration when we change key
parameters in the wind model.  Unless otherwise indicated, parameters
retain fiducial values adopted for the purpose of this paper
(see Table~\ref{paramTable}).  A summary of the variations that we
have explored is shown in Table~\ref{surveyTable}.

\subsubsection{Results}

Key results from our parameter survey are shown in
Figures~\ref{WindVaryCloud} to \ref{WindVaryMBH}, which are described
in detail below.

A very interesting result from this survey is how important embedded
clouds are in driving the wind to new geometries and higher
velocities.  The most dramatic example of the interaction of clouds is
shown in Figure~\ref{WindVaryCloud}, where we vary both the mass of
the clouds and the mass outflow rate of cloud material.  In this
figure, for high cloud mass-outflow rates, we find large changes in
wind structure and kinematics.

The great importance of clouds stems from four major differences
between the clouds and the wind: the low volume filling factor of the
clouds, the lower ionization of the cloud material due to high density
relative to the continuous wind, dust within the clouds, and the fact
that our clouds are not magnetically driven -- they are free to follow
purely radial trajectories through the centrifugally driven wind.  The
low filling factor decreases the opacity of cloud material, thereby
decreasing $t_{\rm cloud}$ (see the definition of $t_{\rm cloud}$ in
eq.~[\ref{tCloudEqn}]).  This low filling factor then leads to an
increase in the line force multiplier (see eq.~[\ref{mLinesDef}] and
Fig.~\ref{forceMultCompare}, which shows the increase in force
multiplier with decrease in $t$) and accelerates the clouds much more
than the surrounding, continuous wind, where the line opacity is much
higher.  The clouds can then drag the wind with them and significantly
alter the wind when the mass outflow fraction of the clouds is high
enough.  As is shown in Figure~\ref{WindVaryCloud}, for $\dot{M}_{\rm
cloud} = 0.5~\dot{M}_{\rm wind}$, the clouds have a very large impact
on the continuous wind's geometry and velocity.  This is due largely
to the large mass outflow rate in the clouds, but also to the lower
opacity of the cloud material, as well as the lower ionization
parameters that exist in the more dense clouds.  The lower ionization
parameters in the clouds mean that the gas is less ionized, yielding
more possible continuum transitions, and resulting in a higher
continuum force multiplier (the increase of the continuum force
multiplier with the ionization parameter, $U$, is also shown in
Figure~\ref{forceMultCompare}).

Dust can also be an important source of opacity and hence radiative
driving for clouds, although for many of the outflows within our
survey, dust cannot survive because its temperature in the clouds
quickly rises above the sublimation temperature.  In general, in
survey, dust only survives in the clouds when there is very
significant shielding of a high-density wind ($N_H \geq 10^{26} {\rm
cm}^{-2}$), a comparatively low central luminosity (of order $0.1
L_{\rm Edd}$), or if the clouds are contained and shielded in a dusty
wind.  In each of these cases, the resultant low luminosity striking
the clouds means that even though dust is present, very little
radiative acceleration results from it.  However, if we move the wind
out to larger radii, as shown in Figure~\ref{WindVaryR}, dust is able
to survive in the clouds and asserts critical importance in modifying
the wind structure: in the case of the $r = 6 \times 10^{19}$~cm
solution, the radiative acceleration due to the continuum in the
clouds overwhelms line driving in the clouds by a factor of $\sim
100$.  This is also the reason behind the increase in velocity for the
$n_0 = 10^6~{\rm cm}^{-3}$ model in Figure~\ref{WindVaryDensity},
where we show the effect of modifying the wind's initial density.
Dust can also survive within the clouds in this model, yielding higher
continuum radiative acceleration, and pushing the wind to higher
velocities than otherwise possible.  Without dust, the clouds within
the $n_0 = 10^7~{\rm cm}^{-3}$ solution achieve similar velocities
only near the disk, but cannot match the increased terminal velocity
in the $n_0 = 10^6~{\rm cm}^{-3}$ outflow.  The $n_0 = 10^7~{\rm
cm}^{-3}$ wind can achieve similar velocities near the disk due to higher
density clouds.  The clouds in this model have higher density because
of the wind's higher density: the wind with greater density requires a
higher magnetic field strength to launch the wind, which confines the
clouds with higher density.  The higher cloud density, in this case,
results in an ionization parameter that is an order of magnitude
lower, allowing normal continuum radiation driving to help drive the
clouds to higher velocities.



	The presence of clouds also explains another effect: close
examination of Figure~\ref{WindVaryDensity} shows an interesting
near-correspondence of velocities for models with $n_0 = 10^7~{\rm
cm}^{-3}$ and $n_0 = 10^9~{\rm cm}^{-3}$ toward the end of those
outflows.  For the lower density ($n_0 = 10^7~{\rm cm}^{-3}$) wind,
clouds push the wind to higher velocities.  As the base wind density
climbs to $10^8~{\rm cm}^{-3}$ and $10^9~{\rm cm}^{-3}$, the clouds
become less and less important, and continuum radiative driving
primarily accelerates the continuous wind.  The continuum driving
dominates the wind briefly at the base of the wind, before the wind is
overionized by the continuum, and then contributes again at larger
distances when the clouds are much further from the central continuum
and the ionization again drops with the decrease in incident flux.
This effect leads to a delayed acceleration of the higher-density
models relative to the lower-density models, whose lower-ionization
clouds are accelerated by continuum radiative driving much closer to
the disk, as can be seen in Figure~\ref{WindVaryDensity}.

The radiative acceleration of the continuous wind itself is also shown
in Figure~\ref{WindVaryShield1e9}, where we present wind models with
varying gas columns in front of the radiatively driven wind (the gas
column shields the wind from too much ionizing radiation, allowing
more efficient radiative acceleration; see MCGV95).  The higher
velocities attained by the radiatively driven winds relative to the
pure-MHD wind are due to continuum radiative driving on the gas in the
continuous wind itself.  The clouds here are relatively unimportant.
But this radiative acceleration is very dependent on density in the
wind.  When we examine the same variation in shielding but with a
lower initial wind density in Figure~\ref{WindVaryShield1e8}, we see
that radiative acceleration of the continuous wind is not important;
the slight variations in the terminal velocity in
Figure~\ref{WindVaryShield1e8} are again due to clouds dragging the
continuous wind as the clouds are accelerated radiatively.

Radiative acceleration can be important in the continuous wind in
other circumstances, however: if we include dust within the continuous
wind, this increases the continuous opacity and therefore the
continuum radiative acceleration.  In the case of
Figure~\ref{WindVaryDust}, where we include dust in the continuous
wind model, the radiative acceleration due to continuum opacity
overwhelms the line driving by a factor of over $100$.  One can also
see the domination of continuum driving in another way: winds of
different density (compare the $n_0 = 10^8$~cm$^{-3}$ to the $n_0 =
10^9$~cm$^{-3}$ run in Fig.~\ref{WindVaryDust}) have extremely similar
structure and velocities when both winds include the same type of
dust. [For both of these models, we have included ISM dust
\citep{MR91,MRN77} in the photoionization and radiative acceleration
models.]  Finally, it is very interesting to note that if we include a
different kind of dust, such as Cloudy's Orion dust \citep[see also
\S\ref{dustExpl}]{Baldwin91}, we see significantly different
velocities (a 30\% change in the terminal velocity of the wind with
Orion dust versus a 20\% change with ISM dust).  This difference may
be of use to researchers trying to gain a better understanding of AGN
dust composition \citep[e.g.,][]{CK01}.

In the remainder of our survey, radiative acceleration plays a key
role in wind acceleration only very near the surface of the accretion
disk, and has a much smaller role in determining the terminal
velocities.  The greatest changes in velocity near the accretion disk
are exhibited for models with small amounts of gas shielding, such as
those shown in Figure~\ref{WindVaryShield1e9}.  In this figure, the
chief culprit behind the large changes to velocity is continuum
driving of the continuous wind.  As the shielding in front of the wind
is increased, the continuum flux drops, and so does the acceleration
near the disk.  Also, it is interesting to note that at these high
densities and with clouds added into the continuous wind, the
velocities are rather insensitive to the incident continuum, and show
very much the same structure with either the MF87 or MCGV95 spectral
energy distribution (see Fig.~\ref{WindVarySpectrum1e9}).  However,
the models do show a slight sensitivity to the incident spectrum for
some values of the shielding column in front of the wind, as also
displayed in Figure~\ref{WindVarySpectrum1e9}: the lower shielding
allows for increased radiative acceleration nearer to the disk
surface, resulting in higher continuum acceleration near the disk.
The MF87 spectrum is slightly more effective in continuum
acceleration, yielding the small difference in velocities between
these two cases.

Large acceleration near the disk is also displayed in outflows
where the wind is launched closer to the central black hole, as in
Figure~\ref{Windr6e17}; this is simply due to the greater amount of
flux at that smaller distance from the black hole.  This flux
primarily powers the acceleration of the wind through a short burst of
continuum driving of the continuous wind as the outflow launches from
the disk, before it is overionized by the intense radiation field.
The higher density winds maintain lower ionization continuous winds
further up in the flow, allowing more continuum acceleration, and
accounting for the slightly larger terminal velocities in those cases.

The acceleration near the disk can also increase if the pure-MHD wind
streamlines are less vertical and more radial (allowing radiative
acceleration to work much more efficiently, since the radial direction
is the direction in which photons are streaming in our model).  This
effect is displayed in Figure~\ref{WindVaryKappa1e9}, where we vary
$\kappa$, the dimensionless ratio of mass to magnetic flux in the
wind. Higher values of $\kappa$ result in streamlines that lift gas
much more slowly from the disk's surface, and therefore are more
radial.  If we also vary $\kappa$ as in Figure~\ref{WindVaryKappa1e9},
but for a lower-density wind, radiative acceleration of the continuous
wind is not as effective, and the acceleration of the wind near the
disk is about an order of magnitude lower (see
Fig.~\ref{WindVaryKappa1e8}).

The above cases all show very high radiative acceleration near the
accretion disk.  An order of magnitude less acceleration near the disk
is shown in many other models (see Figs.~\ref{WindVaryKappa1e8} --
\ref{WindVaryMBH}), although even this decreased force still results
acceleration that is a factor of $\sim 2$ higher than in a pure
centrifugally-driven wind.  This variation of $\sim 2$ is very much
like the the radiative acceleration of the fiducial $n_0 = 10^8~{\rm
cm}^{-3}, N_{\rm H} = 10^{25}~{\rm cm}^{3}$ solution on which these
models are based; the fact that these figures show relatively little
perturbation from that previous model shows that the radiative
acceleration is only weakly dependent on perturbations in these
parameters.  In all but the lowest-density cases, this added
acceleration is fairly constant, even with changes in the central
continuum.  However, the acceleration does drop when the central
luminosity decreases from the fiducial value of $0.5~L_{\rm Edd}$ to
$0.1~L_{\rm Edd}$ (see Fig.~\ref{WindVarySpectrum1e8}). 

In many of the above figures (e.g., Fig.~\ref{WindVarySpectrum1e8})
one may notice a slow oscillation in density with height above the
disk.  The large initial difference in density between the
centrifugally driven wind and the radiatively modified wind is due to
the acceleration of the wind close to the disk: because of mass
conservation, radiative acceleration results in a large drop in the
density compared to that in the pure MHD wind.  Acceleration far from
the disk causes the same effect.  In both cases, we essentially
witness the rise and fall of the ionization parameter, and with it,
the corresponding decrease and increase of radiative acceleration.
The large gas column near the surface of the disk leads to low
ionization parameters there, allowing continuum radiative driving to
dominate, accelerating the wind near the disk.  As the wind climbs,
the shield column drops, the ionization parameter increases, and
radiative acceleration drops.  As this magnetically collimated wind
rises, it gradually climbs further and further away from the disk, and
hence also from the central black hole; as the wind climbs, therefore,
the flux striking the wind drops faster than the wind density, and the
ionization parameter, $U$, again decreases.  The lower ionization
state leads to more lines for radiative acceleration, causing both
line driving and continuum driving to increase, although continuum
driving retains its dominance even far from the central source.


Before concluding this parameter survey, we briefly touch on the
velocity differences predicted by this wind model when the mass of the
central black hole is varied.  These results are shown in
Figure~\ref{WindVaryMBH}, which seems to present a rather paradoxical
picture: as we increase $M_{\bullet}$ and the luminosity increases
(since it is proportional to the Eddington value), the wind's
structure, velocity, and density approach those of the initial,
pure-MHD wind.  In fact, the largest change in the streamlines, and
largest change in the velocity, occur for the lowest mass black hole.
This is due to our assumption that $L \propto L_{\rm Edd}$: for lower
$M_{\bullet}$, this lower luminosity allows a lower ionization
parameter in the continuous wind, which yields more continuum opacity
in the wind, and hence larger continuum acceleration.  Also, for the
the case of $M_{\bullet} = 10^7 M_{\sun}$, the luminosity is low
enough for dust to survive in the clouds, which further aids in the
acceleration, although the continuum driving of the wind is still the
overriding source of acceleration.


\section{Conclusions \& Future Directions}\label{Conclusions}

We have presented a self-consistent, semianalytic, steady-state
disk-wind model that combines magnetic and radiative accelerations and
includes two gas phases: a magnetized continuous wind and embedded
diamagnetic clouds.  The continuous wind is driven centrifugally and,
after rising above the disk surface, becomes subject to radiative
acceleration by the central continuum source.  The clouds are uplifted
from the disk surface by the ram pressure of the continuous wind,
which also confines them by its (largely magnetic) pressure, and they
are subsequently pushed by the radiation pressure force of the central
continuum.  We calculate the radiative acceleration on both of those
components from first principles, using detailed photoionization
simulations to determine the ionization state of the gas and the
continuum incident on that gas.  Since the model includes magnetic
launching, the outflow automatically incorporates a shielding column
of gas that can attenuate the central continuum and allow efficient
radiative acceleration in the outer parts of the wind.  This addresses
the problem of how to launch a shielding column in areas of the wind
where radiative acceleration is not effective, because of the high
degree of ionization of the gas.

We have used this model to illustrate the dependence of a two-phase
magnetically and radiatively driven wind on various outflow
parameters, and to demonstrate the care with which one must treat
radiative acceleration calculations.  We find that {\it the wind
structure is significantly altered} when we include clouds that are
also radiatively accelerated: their low filling factor, low ionization
parameters, and the possibility of dust in the clouds all allow
radiative acceleration when the continuous wind is too highly ionized
for radiation to be an effective driving source.  Not unexpectedly,
clouds have the largest impact when their mass outflow rate is
comparable to that of the wind itself.  We have also determined that
including dust in the continuous outflow can have a large impact on
wind geometry and kinematics because of the increase in continuum
opacity.  Dust has a large enough impact on wind structure that even
changing the type of dust results in significant changes in the
outflow.  Finally, we have found that higher-density winds with low
shielding can be influenced heavily by radiative acceleration.  In
this case, radiative acceleration is not acting through the clouds but
instead comes from the continuum driving in the continuous wind
itself; large changes in the wind occur with dust present in the wind,
but even the gas continuum opacity can result in large radiative
acceleration of the continuous wind.  This is a recurring theme in our
results: continuum driving by and large plays a more important role in
continuous wind acceleration than line driving.  Line driving, on the
other hand, is more important in the clouds.

We have also found that, in the context of our particular cloud model,
much of the spectral dependence of radiative acceleration is largely
erased, with different central continua producing very similar
perturbations to the wind's structure and dynamics.  Furthermore, we
showed that variations in the mass of the central black hole and in
the launching radius of the wind do not greatly modify the initial
centrifugally-driven wind's structure or kinematics, although it is
important to remember that changing both of these parameters affects
the Keplerian velocity at the base of the wind, and hence changes the
magnitude of the wind velocity.

Finally, we established that the velocity structure near the surface
of the disk is very sensitive to radiative acceleration, which may be
an important key to observationally distinguishing between pure
MHD-driven winds and radiatively driven winds.  While both accelerate
rapidly from the disk, radiation pressure causes much greater initial
acceleration as well as much larger drops in the density as the wind
lifts off from the surface of the disk.  This effect would be much
more noticeable in AGNs with low shielding columns, where radiative
acceleration is much stronger near the surface of the disk.  As a
result of mass conservation, the variations in density with altitude
above the disk closely mirror those of the velocity, and may also be
valuable tracers to discriminate between wind acceleration processes.
This may hint that if Broad Emission Lines are formed near the disk
surface, observations of those lines may be capable of distinguishing
between different acceleration processes.  It is important to note
that we have not yet included the radiative acceleration due to
intrinsic or reprocessed flux from the accretion disk itself as in
MCGV95 and \citet{Pro00}.  Including this effect may yield even
stronger acceleration from the surface of the disk.

Our next step will be to fit this model to observations in an attempt
to directly quantify the processes that act within AGN outflows.  To
complete this, it will be important to widen our parameter survey to
include, for example, other incident spectra and dust compositions
\citep[to examine, for instance, the dust distribution found
in][]{CK01}.  We are currently developing a Monte Carlo
module that will allow both line emission and absorption predictions
for a range of inclination angles.  At the same time, we also hope to
include the effect of radiation acceleration due to accretion disk
flux and pursue observational signatures of embedded condensations
that are not magnetically confined by the continuous wind
\citep[e.g.,][]{Bot01}.  In addition, we can learn a great deal about
the geometry of AGN outflows from polarization measurements
\citep[e.g.,][]{Corbett00, Smith02}.  The calculation of both line and
continuum polarization will also be included in our Monte Carlo
program.  Finally, we note that this program is in no way restricted
to considering AGN outflows alone; one can easily change the model's
input parameters to study systems such as high-luminosity young
stellar objects and cataclysmic variables, where similar outflows are
observed.


\section{Acknowledgements}
I gratefully acknowledge the constant support of Dr. Arieh K\"onigl,
my advisor.  He was an invaluable source of guidance when this project
was taking shape, as well as when I was testing the various modules,
trying to understand their behavior/misbehavior.  Special thanks,
also, to Dr. John Kartje, whose early work on AGN wind models helped
form the basis for this work.  I am also thankful for the help of
Drs. Lewis Hobbs, Bob Rosner, and Don York for valuable questions and
discussion, as well as discussions with Drs. Steve Kraemer and Jack
Gabel.  And finally, none of this work would have been possible
without the generous support of NASA's ATP program, in this case, in
the guise of grant NAG5-9063.

\appendix

\section{Derivation of the Self-Similar Centrifugal Wind Equations}\label{selfSimWindEqns}

In this appendix, we sketch the rederivation of our system of
self-similar wind equations for the magneto-centrifugal wind
structure.  The equations utilized in this calculation advance upon
those presented in BP82 and KK94: we consider not only a wind with
arbitrary density power-law index, $b$, as in KK94, but also rederive
the equations without the constraint of energy conservation.  Since
the radiation field continually inputs energy into the outflow, this
is a very important modification, and was not fully considered within
the derivation presented in KK94.

First, we work under the assumptions of stationary, axisymmetric,
ideal, cold magnetohydrodynamic flow in cylindrical coordinates
$(r,\phi,z)$.  Our system of equations is based upon both the radial
and vertical momentum equations:
\begin{eqnarray}
v_r \frac{\partial v_r}{\partial r} + v_z \frac{\partial v_r}{\partial
z} - \frac{v_{\phi}^2}{r} & = & - \rho \frac{\partial \Phi}{\partial r} - 
\frac{B_z}{4 \pi} \left( \frac{\partial B_z}{\partial r} -
\frac{\partial B_r}{\partial z} \right) - \frac{B_{\phi}}{4 \pi r}
\frac{\partial (r B_{\phi})}{\partial r} \\
\rho (\mathbf{v} \cdot \mathbf{\nabla}) v_z & = & -\rho \frac{\partial
\Phi}{\partial z} - \frac{1}{8 \pi} \frac{\partial B^2}{\partial z} +
\frac{1}{4 \pi} (\mathbf{B} \cdot \mathbf{\nabla}) B_z, \\
\end{eqnarray}
where $\mathbf{v}$ is the fluid velocity, and $\mathbf{B}$ is the
magnetic field strength.  We neglect the thermal term, as we work in
the zero-temperature limit.  $\Phi$ is the effective gravitational
potential, defined as
\begin{eqnarray}
\Phi = - [1 - \Gamma(\theta)] \frac{GM_{\bullet}}{(r^2 + z^2)^{1/2}},
\end{eqnarray}
where $M_{\bullet}$ is the mass of the central black hole, and
$\Gamma(\theta)$ gives the ratio radiative acceleration relative to
the local gravitational field strength (see eq.~\ref{gammaDef}).

We can solve these equations by first relating the flow velocity to
the magnetic field via \citep[see, e.g., ][]{Chandra56, Mestel61}:
\begin{eqnarray}
\mathbf{v}(\mathbf{r}) = \frac{k \mathbf{B}(\mathbf{r})}{4 \pi
\rho(\mathbf{r})} + (\mathbf{\omega}(\mathbf{r}) \times \mathbf{r}), 
\end{eqnarray} 
and stipulating mass conservation, 
\begin{eqnarray}
\mathbf{\nabla} \cdot (\rho \mathbf{v}) = 0, 
\end{eqnarray}
where $k/4\pi$ is the ratio of mass flux to magnetic flux,
and $\mathbf{\omega}(\mathbf{r})$ and $\rho(\mathbf{r})$ are the angular
velocity and mass density of the gas in the flow.  Both the $\omega$
and $k$ are conserved along magnetic field lines.

This system would normally have two integrals of motion: the specific
energy and the specific angular momentum.  However, since energy is
continually added to the outflow, the specific energy is constant, and
we can use only the specific angular momentum
\begin{eqnarray}
l = r v_{\phi} - \frac{r B_{\phi}}{k}.
\end{eqnarray}

We then impose self-similarity on this system by specifying
\begin{eqnarray}
\mathbf{r} & = & [r_0 \xi(\chi), \phi, r_0 \chi] \\
\mathbf{v} & = & [\xi'(\chi) f(\chi), g(\chi), f(\chi)]v_{K0},
\end{eqnarray}
where $v_{K0}$ is the Keplerian velocity at the base of the outflow,
$v_{K0} = (GM_{\bullet}/r)^{1/2}$, and the prime indicates
differentiation with respect to $\chi$.  At the same time, we can
re-express our constants in dimensionless form:
\begin{eqnarray}
\lambda & \equiv & \frac{l}{(GM_{\bullet} r_0)^{1/2}} \\
\kappa & \equiv & \frac{k (1+{\xi'}_0^2)^{1/2}}{B_{p0}} v_{K0},
\end{eqnarray}
where $B_{p0}$ is the initial poloidal magnetic field strength at the
base of the wind.

As in KK94, we allow for a general power-law scaling of the density
and magnetic field along the disk's surface:
\begin{eqnarray}
\rho_0 & \propto & r_0^{-b} \\
B_0 & \propto & r_0^{-(b+1)/2}.
\end{eqnarray}

With this self-similar specification, the radial and vertical momentum
equations become, after some simplification:
\begin{eqnarray}
\frac{f \xi' m'}{\kappa \xi J} - \frac{f^2 \xi'}{\xi J} +
\xi'' f^2 - \frac{(\lambda m - \xi^2)^2}{\xi^3 (m-1)^2} & = & 
- \xi [1 - \Gamma(\theta)] S^3 - \frac{f}{\kappa \xi J^2} \left( 
\frac{-(1+\xi'^2)(b+1)}{2} + \nonumber \right. \\
& & \left. \frac{(\chi + \xi \xi')\xi'}{\xi} -
\frac{\xi''}{J S^2} \right) - \frac{\kappa f}{\xi} \frac{(\lambda -
\xi^2)}{(m-1)} \nonumber \\ & & \left[ \frac{(\lambda -
\xi^2)}{(m-1)}\frac{(-b+1)}{2} + 
\chi \left( \frac{2 \xi \xi'}{(m-1)} + \frac{(\lambda - \xi^2)m'}
{(m-1)^2} \right) \right] \\
\frac{f}{\kappa \xi J} (m' - f \kappa \xi' J + f \kappa \xi
\chi \xi'')  & = &
 -[1 - \Gamma(\theta)]\chi S^3 + \frac{f \xi'}{\kappa \xi J^2}
 \left( \frac{-(1+\xi'^2)(b+1)}{2} + \nonumber \right. \\ & & \left. \frac{(\chi + \xi \xi')\xi'}{\xi} -
\frac{\xi''(\chi^2 + \xi^2)}{J} \right) - \nonumber \\
& & \xi' \kappa f (\lambda - \xi^2)
\left(\frac{(b+1)(\lambda-\xi^2) - 2(\lambda+\xi^2)}{2 \xi (m-1)^2}
\right) + \nonumber \\
& & \frac{(\lambda - \xi^2)^2 m' \kappa f}{(m-1)^3}.
\end{eqnarray}
In the above, the following definitions are in play:
\begin{eqnarray}
m & = & f \kappa \xi J = \frac{4 \pi \rho v_p^2}{B_p^2}  =  {\rm
square~of~the~Alfven~Mach~number} \\
\Gamma(\theta) & = & \frac{a_{radiative}}{g}\\
k/4 \pi & = & {\rm ratio~of~mass~flux~to~magnetic~flux~(constant)} \\
\kappa & = & \frac{k (1 - {\xi'}_0^2)^{\frac{1}{2}} v_{k0}}{B_{p0}}  =
{\rm dimensionless~ratio~of~mass~flux~to~magnetic~flux}\\
l & = & r v_{\phi} - \frac{r B_{\phi}}{k} = {\rm specific~angular~momentum}\\
\lambda & = & \frac{l}{(G M r_0)^{\frac{1}{2}}} = {\rm normalized~angular~momentum}\\
J & = & \xi - \chi \xi' \\
S & = & 1/\sqrt{\xi^2 + \chi^2} \\
\rho & \propto & R^{-b}. \\
\end{eqnarray}
These two relations define the differential equations for $m'$ and
$\xi''$, which are, respectively, the spatial gradient in the poloidal
Alfv\'en mach number (gradient with respect to height, $\chi$) and the
(cylindrical) radial velocity gradient (again with respect to $\chi$).

One can see, from close inspection of the above equations, that many
of the terms have a denominator of $(m - 1)$, showing that when the
gas crosses the Alfv\'en point (defined to be where $m = 1$), the
equations become singular.  The point $m = 1$ is therefore a critical
point in the flow, where the downstream gas accelerates beyond the
speed where upstream Alfv\'en waves can communicate with the rest of
the wind.  We can rewrite and solve the $m'$ equation for the value of
$m'$ at the Alfv\'en point, with the result being
\begin{eqnarray}
m'_{A} & = & 2 \xi J [-8 \chi \kappa^2 \lambda m' \xi' J^3 + 4 (1+b)
\kappa^2 \lambda \xi'^2 J^2 (\chi + \xi \xi') + m'^2 (\chi + \xi \xi')
(-2 \kappa^2 \lambda^2 S + \nonumber\\ & & (1+b) + 2 \kappa^2 \lambda^3 - 
4 \chi
\kappa^2 \lambda^{\frac{3}{2}}(\lambda - S) \xi' + ( (1+b) + 2 \chi^2
\kappa^2 \lambda (\lambda - S))\xi'^2 + \nonumber \\ & & 2 \kappa^2 \lambda
\Gamma(\theta) J^2)] / \left[ 4 \xi J \left( \frac{4 \kappa^2 \lambda \xi'^2
J^2}{S^2} + m'^2 (\chi + \xi \xi')^2 \right) \right]. \label{mpAeqn}
\end{eqnarray}
We then use this constraint to our advantage, and start the integral
at the Alfv\'en point with the value of $m'_A$ given by
Equation~\ref{mpAeqn}.

As covered in the main text, we then integrate these equations using a
``shooting algorithm'' to integrate both from the critical point and
the disk surface, towards an intermediate point.  Matching these two
integrals at this common, central point allows us to solve for the
three free parameters in our system, $\xi'_0$, $\xi'_A$, and
$\chi_A$.  

\section{Cloud Calculation Initial Conditions}\label{cloudInit}

For almost all cloud simulations, we use Cloudy to determine the
internal cloud parameters.  However, at the very base of the wind,
Cloudy is very often unable to simulate the shielded, high density
clouds, as almost all of the carbon within the clouds becomes
molecular.  Therefore, we must define alternative approximations to
the initial cloud parameters, which we do as follows:

\begin{description}
\item[$T_{Cloud,init}$:] In our MHD wind models, we have already made
the assumption that thermal driving (gas pressure) is unimportant.
Physically, this means that the initial velocity of the wind (where we
start tracking it) must be greater than or of order the local sound
speed; otherwise, thermal effects can be communicated within the wind,
and would then not be negligible, as we have assumed.  We can use this
assumption to estimate the temperature in the clouds.  Basically, if
we assume that the continuous wind velocity is of order of the sound
speed in the wind, we can use that known initial wind velocity to
estimate the temperature in the wind.  We then approximate that the
cloud temperature is of order the wind temperature at the disk's
surface, which should be reasonable since the clouds form there.
Thus, we can write
\begin{eqnarray}
v_{wind,init} & = & c_s =\sqrt{\frac{k_B T_{wind,init}}{\mu m_p}} = \sqrt{\frac{k_B T_{cloud,init}}{\mu m_p}} \\
\Rightarrow T_{cloud,init} & = & \frac{\mu m_p v_{wind,init}^2}{k_B},
\end{eqnarray}
where $c_s$ is the speed of sound in the wind and $\mu$ is the mean
particle mass in units of the mass of the proton, $m_p$.  For gas near
the disk, we set $\mu = 1$ for simplicity.


\item[$n_{cloud,init}$:] Having some idea of the initial temperature
in the clouds, we now hypothesize (as we will throughout the model)
that the clouds are in pressure equilibrium with the wind.  For our
model, the wind's pressure is dominated by the magnetic pressure.  We
are assuming that the clouds are diamagnetic so that $P_{\rm
mag}~\ll~P_{\rm thermal}$; this also means that the clouds are not
guided or accelerated by the magnetic field in the continuous wind.
Since we already know the wind's magnetic pressure from the MHD
calculations, we write
\begin{eqnarray}
P_{wind} & \approx & P_{wind,mag} = P_{cloud} \\
\Rightarrow n_{cloud,init} & = & \frac{P_{wind,mag}}{k_B T_{cloud,init}}.
\end{eqnarray}

\item[$R_{cloud,init},M_{cloud,init}$:] We solve for both of these
simultaneously, given that each individual cloud mass,
$M_{cloud,init}$, is some user-specified fraction of the maximum
individual cloud mass that the wind could push up from of the disk by
ram pressure ($M = \rho V$ where $\rho$ is the mass density in the
cloud and $V = \frac{4}{3}\pi R^3$ is the volume of the cloud; we
assume spherical clouds for simplicity but admit that diamagnetic
clouds may certainly be forced into other shapes).  The equation for
the maximum cloud mass is given by setting the ram pressure of the
wind equal to the gravitational force on the cloud mass.  In this
case, for a mass rising up above the disk, the gravitational force to
overcome is the tidal force, which is a fraction $\frac{h}{r}$ of the
gravitational force due to the central source, where $h$ is the scale
height of the disk.  Therefore, we write \citep[as in][]{KKE99}
\begin{eqnarray}
\frac{G M_{\bullet} M_{cloud,max}}{r_{Launch}^2} \frac{h}{r} & \approx &
C_F \pi R_{cloud,init}^2 \rho_{wind,init} v_{wind,init}^2 \\
\Rightarrow M_{cloud,max} & \approx & \frac{C_F \pi
R_{cloud,init}^2 \rho_{wind,init} v_{wind,init}^2 r_{Launch}^3}{G
M_{\bullet} h}.
\end{eqnarray}
But we also know that
\begin{eqnarray}
M_{cloud} & = & \delta M_{cloud,max} = \frac{4}{3} \pi
R_{cloud,init}^3 n_{cloud,init} m_p \\
\Rightarrow M_{cloud,max} & = & \frac{4}{3 \delta} \pi
R_{cloud,init}^3 n_{cloud,init} m_p,
\end{eqnarray}
where the user specifies $\delta$.
Setting the two expressions for $M_{cloud,max}$ equal, we can solve
for $R_{cloud,init}$
\begin{eqnarray}
R_{cloud,init} & \approx & \frac{3 \delta}{4} \frac{C_F
v_{wind,init}^2 r_{Launch}^3}{G M_{\bullet} h} \frac{n_{wind,init}}{n_{cloud,init}}.
\end{eqnarray}
A useful approximation for thin accretion disks is that the ratio
$\frac{h}{r} \approx \frac{c_s}{v_k}$ where $c_s$ is the sound speed
and $v_k$ is the Keplerian velocity \citep[we note that $h$ obtained in
this way is an upper limit if magnetic squeezing of the disk is
important; see][]{WK93}.  As already mentioned, very near
the disk we assume that the wind is transonic, hence $v_{p,0} \geq c_s$.
We therefore set $c_s \approx v_{p,0}$, since we know $v_{p,0}$ from
the MHD calculations.  Substituting this into our equation, and also
setting $v_{wind,init}$ equal to the velocity at the base of the wind,
\begin{eqnarray}
R_{cloud,init} & \approx & \frac{3 \delta}{4} \frac{C_F
v_{p,0} r_{Launch}^2 v_k}{G M_{\bullet}} 
\frac{n_{wind,init}}{n_{cloud,init}}. \\
\end{eqnarray}
We use this equation to calculate the initial radius of the clouds
that are launched into the wind.

\item[$n_{cloud,ens}$:] When we compute the interaction of the clouds
with the wind, we need to know the ensemble density of clouds (i.e.,
the number density of clouds within the wind).  We estimate
$n_{cloud,ens}$ using the equation
\begin{eqnarray}
\dot{M}_{cloud} & = & M_{cloud} n_{cloud,ens} v_{cloud} A \\
\Rightarrow n_{cloud,ens} & = & \frac{\dot{M}_{cloud}}{M_{cloud}
v_{cloud} A},
\end{eqnarray}
where $\dot{M}_{cloud}$ is the mass outflow rate due to clouds and $A$
is the cross-sectional area of the outflow.  At this stage of the
flow, near the disk, we approximate $A = \pi (r_{outer}^2 -
r_{inner}^2)$.  Next, we define a user-specified parameter
\begin{equation}
\eta \equiv \frac{\dot{M}_{cloud}}{\dot{M}_{wind}},
\end{equation}
which allows us to write
\begin{eqnarray}
\Rightarrow n_{cloud,ens} & = & \frac{\eta \dot{M}_{wind}}{M_{cloud}
v_{cloud} A}.
\end{eqnarray}

\end{description}

The most important of the above parameters is the cloud mass, which
requires an evaluation of all of the above parameters except for
$n_{cloud,ens}$.  Above the accretion disk, pressure equilibrium with
the magnetic wind (determined through Cloudy) will set the density of
a cloud, and since we already know the cloud mass, we can then
calculate the radius of the cloud.  This is necessary for computing
the drag forces on the cloud (which scale as $R_{cloud}^2$).

Once all of the initial parameters have been found, we proceed to
simulate clouds at any point in the flow.

\end{document}